\documentclass[11pt]{article}
\usepackage{fullpage,epsf,amssymb,amsthm,amsfonts,amsmath,latexsym,array,graphicx,extarrows,mathtools,mathstyle,mathrsfs,slashed,subcaption,amsbsy,csquotes,accents}
\usepackage[normalem]{ulem}
\usepackage{authblk}
\usepackage{cite}
\usepackage{color}    
\usepackage{float}
\usepackage{hyperref}
\usepackage[export]{adjustbox}
\usepackage{hhline}
\usepackage{enumitem}

\numberwithin{equation}{section}

\title{\bf{Towards a consistent perturbation theory at finite temperature}} 

\author[1]{Passant Ali}
\author[1]{Peter Lowdon}
\author[1]{Owe Philipsen}

\affil[1]{{\scriptsize Institut f\"{u}r Theoretische Physik, Goethe-Universit\"{a}t, Max-von-Laue-Str. 1,  60438 Frankfurt am Main, Germany}}

\date{}
\begin{document}

\maketitle

\begin{abstract}
\noindent
The standard approach to perturbation theory for finite-temperature quantum field theories has several issues, including the appearance of ill-defined on-shell contributions in the real-time formulation, and infrared divergences in massless theories. Earlier studies indicate that these issues all stem from the inconsistent thermal generalisation of the Gell-Mann-Low relation, which forms the foundation of perturbation theory in vacuum. This inconsistency arises from the use of free scattering states in the relation, which are known not to exist in interacting thermal theories. In this work, we propose a generalisation of the Gell-Mann-Low relation for scalar theories based on non-perturbative spectral insights, namely that finite-temperature scattering states can be described by damped but stable particle-like excitations, so-called thermoparticles. The perturbative expansion of this generalised relation gives rise to contributions with exactly the same topology as the standard finite-temperature approach, except that now the propagators appearing in this expansion are not those of a free field but of thermoparticles, which depend on the dynamics of the theory. We demonstrate that thermoparticle perturbation theory resolves the known problems of the standard approach. Furthermore, by comparing imaginary-time calculations at two-loop level with numerical lattice simulations of two-point correlation functions in massive $\phi^{4}$ theory, we explicitly show that this framework gives rise to precise predictions, as in the vacuum case, in stark contrast to the standard approach. 
\end{abstract}

\newpage

\section{Introduction}
\label{intro} 

Perturbation theory provides a powerful approach to extract dynamical information about quantum field theories (QFTs) in weakly interacting regimes. For QFTs at zero temperature the theoretical foundation for a perturbative evaluation is provided by the Gell-Mann-Low (GML) relation~\cite{Gell-Mann:1951ooy}, which can be rigorously proven in certain cases\footnote{See Refs.~\cite{Landsman:1988ta,Haag:1992hx} and the references within for more details regarding the derivation of the GML relation.}. For scalar QFTs with a single field $\phi(x)$ it has the form
\begin{align}
\langle \Omega|T\!\left\{ \phi(x_{1})\cdots \phi(x_{n}) \right\}|\Omega\rangle = \frac{\langle \Omega_{0}|T\!\left\{ \phi_{0}(x_{1})\cdots \phi_{0}(x_{n}) e^{i \int d^{4}z \, \mathcal{L}_{I}\left[\phi_{0}(z)\right]} \right\} | \Omega_{0} \rangle }{ \langle \Omega_{0}|T\!\left\{ e^{i \int d^{4}z \, \mathcal{L}_{I}\left[\phi_{0}(z)\right]} \right\} | \Omega_{0}\rangle},   
\label{GML_0}
\end{align}
where $\mathcal{L}_{I}$ is the interacting part of the Lagrangian density, and $T\{\cdots \}$ indicates a time-ordering of the field operators. Equation~\eqref{GML_0} provides a connection between the time-ordered correlation functions of the full interacting theory with vacuum state $|\Omega\rangle$, to those of the free field $\phi_{0}(x)$ evaluated in the non-interacting theory with ground state $|\Omega_{0}\rangle$. The GML relation can therefore be used to compute information about the interacting theory using free-field correlation functions. In particular, since $\mathcal{L}_{I}$ is proportional to the coupling $g$, the exponential can be expanded as a power series in $g$. Using Wick's theorem, the terms to any order in $g$ can then be decomposed into sums of products of free-field propagators $\langle \Omega_{0}|T\left\{ \phi_{0}(x)\phi_{0}(y)\right\} | \Omega_{0} \rangle$. Since products of the position-space propagator can be expressed in terms of convolutions of the corresponding momentum-space propagator
\begin{align}
\widetilde{\tau}_{(0)}(p) = \frac{i}{p^{2}-m^{2}+i \epsilon},
\label{free_prop_T0}
\end{align}
this gives rise to the standard bare perturbation theory~\cite{Peskin:1995ev}, where each term of the series is represented by Feynman diagrams with lines corresponding to the propagator in Eq.~\eqref{free_prop_T0}. A crucial aspect for the success of the perturbative treatment is that the free-field propagators appearing in the expansion have the same analytic structure as the full propagators representing the physical asymptotic states created by the fields $\phi_{\scriptscriptstyle \text{in}/\text{out}}(x)$ before and after a scattering process, and hence one has the relation
\begin{align}
\lim_{x_{0},y_{0} \rightarrow  \infty} \langle \Omega |T\!\left\{ \phi(x)\phi(y)\right\} | \Omega \rangle= Z \, \langle \Omega |T\!\left\{ \phi_{\scriptscriptstyle \text{in}/\text{out}}(x)\phi_{\scriptscriptstyle \text{in}/\text{out}}(y)\right\} | \Omega \rangle= Z \! \int \! \frac{d^4p}{(2\pi)^4} \frac{i \, e^{-ip(x-y)}}{p^{2}-m_{\scriptscriptstyle \text{phys}}^{2}+i \epsilon},
\label{full_prop_T0}   
\end{align}
where $m_{\text{phys}}$ is the physical mass of the scattering state. The effect of the perturbative iteration is to renormalise the free fields by $\sqrt{Z}$ and the bare mass $m$ of the free theory to the physical mass $m_{\text{phys}}$. \\

\noindent
For QFTs defined at non-vanishing temperatures $T=1/\beta >0$ several well-known complications appear in the perturbative treatment, including the existence of on-shell singularities and infrared divergences~\cite{Landsman:1986uw}. It has been noted long ago that an expansion in terms of free-field propagators is questionable at finite temperature, since in thermal equilibrium the medium is everywhere-present at all times, and hence there are no asymptotic states corresponding to free particles~\cite{Landsman:1988ta}. Instead, the true asymptotic states must be damped due to medium effects, and thus contain information about the full dynamics of the theory~\cite{Bros:2001zs}. In this study we demonstrate by explicit calculation that the complications encountered in the standard perturbative approach can be avoided by taking these constraints into account in the generalisation of the GML relation in Eq.~\eqref{GML_0}. In Sec.~\ref{sec:PT} we review the standard finite-temperature perturbative approach, focussing on the well-known problems that arise. We propose \textit{thermoparticle perturbation theory} as a new framework in Sec.~\ref{TP_PT}, and explain how it can provide a resolution to these problems. In Sec.~\ref{lattice_test} we test this framework in massive $\phi^{4}$ theory by computing perturbative predictions of correlation functions and comparing with numerical lattice simulations. In contrast to the standard approach, we find that this framework gives rise to percent-level precision predictions which also display consistent convergence properties. Finally, in Sec.~\ref{concl} we summarise our main findings and discuss their implications.

\section{Standard finite-temperature perturbation theory}
\label{sec:PT}

\subsection{The real-time formalism}

In the standard approach to real-time finite-temperature perturbation theory~\cite{Landsman:1986uw,Niemi:1983nf,Bellac:2011kqa} the $T=0$ Gell-Mann-Low relation in Eq.~\eqref{GML_0} is generalised by changing the vacuum expectation value to a thermal one,
\begin{align}
\langle \Omega_{\beta}|T\!\left\{ \phi(x_{1})\cdots \phi(x_{n}) \right\}|\Omega_{\beta}\rangle = \frac{ \text{Tr}\left( e^{-\beta H} \, T_{C}\!\left\{ \phi_{0}(x_{1})\cdots \phi_{0}(x_{n}) e^{i \int_{C} d^{4}z \, \mathcal{L}_{I}\left[\phi_{0}(z)\right]} \right\} \right) }{ \text{Tr}\left(e^{-\beta H} \, T_{C}\!\left\{ e^{i \int_{C} d^{4}z \, \mathcal{L}_{I}\left[\phi_{0}(z)\right]} \right\} \right)}, 
\label{GML_Tv1:1}
\end{align}
where $|\Omega_{\beta}\rangle$ denotes the thermal equilibrium state of the full interacting theory. The trace is taken over the space of states generated by the non-interacting asymptotic fields $\phi_{0}(x)$, and the Lagrangian density $\mathcal{L}_{I}$ is integrated along the so-called Schwinger-Keldysh contour $C=C_{+} \cup C_{-}$, which traverses the real axis in both the forward $C_{+}$ and backward $C_{-}$ directions~\cite{Das:1997gg},
\begin{align}
\int_{C} d^{4}z = \left( \int_{-\infty}^{\infty}  dz_{+}^{0} - \int_{-\infty}^{\infty}  dz_{-}^{0} \right)\int d^{3}z,
\label{SK_int}
\end{align}
where the negative sign comes from the fact that $C_{-}$ is traversed in the opposite direction to $C_{+}$. The time ordering of the field operators $T_{C}\{\cdots\}$ is now taken with respect to the contour to which the fields are restricted~\cite{Das:1997gg}. By interpreting the density operator $e^{-\beta H}$ as an evolution operator in imaginary time, Eq.~\eqref{GML_Tv1:1} can be expressed as
\begin{align}
\frac{ \text{Tr}\left( e^{-\beta H_{0}} \, T_{C}\!\left\{ \phi_{0}(x_{1})\cdots \phi_{0}(x_{n}) e^{i \int_{C \cup C_{v}} \mathcal{L}_{I}\left[\phi_{0}\right]} \right\} \right) }{ \text{Tr}\left(e^{-\beta H_{0}} \, T_{C}\!\left\{ e^{i \int_{C \cup C_{v}} \mathcal{L}_{I}\left[\phi_{0}\right]} \right\} \right)}, 
\label{GML_Tv1:2}
\end{align}
where $C_{v}$ is a vertical contour tranversing from zero to $-i\beta$ in the imaginary-time direction, and $H_{0}$ is the free Hamiltonian operator. With this representation, thermal correlation functions are comprised of multiple components which correspond to the evolution of the fields along different branches of the contour. It is argued that only the integration along $C$ gives a non-trivial contribution~\cite{Bellac:2011kqa}, and hence correlation functions depend solely on whether the fields evolve along $C_{+}$ or $C_{-}$. These restrictions can be expressed as separate fields $\phi^{+}$ and $\phi^{-}$, leading to a doubling of the degrees of freedom. As the fields in Eq.~\eqref{GML_Tv1:2} are non-interacting, a perturbative expansion can be performed analogously to the $T=0$ case by expanding the exponentials in powers of the coupling, and applying a thermal generalisation of Wick's theorem to the resulting correlation functions~\cite{Bellac:2011kqa}. There are four thermal propagators $\tau^{ab}_{(0)}(x-y)$ ($a,b=+,-$) that can contribute due to the different contractions between the $\phi^{+}$ and $\phi^{-}$ fields~\cite{Bellac:2011kqa}
\begin{align}
\begin{array}{ll}
\tau^{++}_{(0)}(x-y) = \langle T\!\left\{\phi^{+}(x)\phi^{+}(y)\right\}\rangle_{\beta},   & \tau^{--}_{(0)}(x-y) = \langle T^{*}\!\left\{\phi^{-}(x)\phi^{-}(y)\right\}\rangle_{\beta}, \\
\tau^{-+}_{(0)}(x-y) = \langle \phi^{-}(x)\phi^{+}(y)\rangle_{\beta},  & \tau^{+-}_{(0)}(x-y) = \langle \phi^{-}(y)\phi^{+}(x)\rangle_{\beta},
\end{array}
\end{align}
where $T^{*}\{\cdots\}$ denotes anti-time ordering, and $\langle \cdot \rangle_{\beta}$ is the thermal expectation value. Along the forward real-time branch $C_{+}$ the propagator $\tau_{(0)}^{++}(x)$ has the momentum space form 
\begin{align}
\widetilde{\tau}_{(0)}^{++}(p) = \frac{i}{p^{2}-m^{2}+i \epsilon} +  \frac{2\pi \delta(p^{2}-m^{2})}{e^{\beta |p_{0}|}-1}.
\label{free_prop}
\end{align}
In the $T\rightarrow 0$ limit this reduces to Eq.~\eqref{free_prop_T0}, as expected. Along the other branches the momentum space propagtors have the structure~\cite{Bellac:2011kqa}
\begin{align}
&\widetilde{\tau}_{(0)}^{+-}(p) = \left[\theta(-p_{0}) + \frac{1}{e^{\beta |p_{0}|}-1}  \right] 2\pi \delta(p^{2}-m^{2}), \nonumber\\
&\widetilde{\tau}_{(0)}^{-+}(p) = \left[\theta(p_{0}) + \frac{1}{e^{\beta |p_{0}|}-1}  \right] 2\pi \delta(p^{2}-m^{2}), \nonumber \\
&\widetilde{\tau}_{(0)}^{--}(p) = \left[\widetilde{\tau}_{\beta}^{++}(p)\right]^{*} = -\frac{i}{p^{2}-m^{2}-i \epsilon} +  \frac{2\pi \delta(p^{2}-m^{2})}{e^{\beta |p_{0}|}-1}.
\label{other_prop}
\end{align}
The fact that the propagator in Eq.~\eqref{free_prop} contains both a temperature-independent and dependent part is often interpreted as arising from a physical decomposition of the system into a quantum component and a heat bath: the first term representing the exchange of a virtual particle, and the second term the contribution of an on-shell excitation from the medium~\cite{Das:1997gg}.

\subsection{Problems with the standard approach}
\label{probs}

Two of the more fundamental problems that occur in the standard approach to perturbation theory at finite temperature are on-shell singularities, and the presence of infrared divergences in massless theories.

\begin{enumerate}[leftmargin=*]

\item {\bf{Ill-defined on-shell components}:} Working in momentum space, the expansion of the full propagator to any non-trivial loop order necessarily involves products of the basic thermal propagators in Eqs.~\eqref{free_prop} and~\eqref{other_prop}. For example, consider the one-loop time-ordered propagator $\widetilde{\tau}_{(1)}^{++}(p)$ in massive $\phi^{4}$ theory, which due to the doubling of the fields has the form~\cite{Das:1997gg}
\begin{align}
\widetilde{\tau}_{(1)}^{++}(p) = \widetilde{\tau}_{(0)}^{++}(p) + i \Delta m_{+}^{2} \left(\widetilde{\tau}_{(0)}^{+-}(p)\widetilde{\tau}_{(0)}^{-+}(p)-\widetilde{\tau}_{(0)}^{++}(p)^{2} \right),  
\label{1-loop++}
\end{align}
where $\Delta m_{+}^{2}$ is the thermal bubble contribution
\begin{align}
\Delta m_{+}^{2} = \frac{\lambda}{2} \int \frac{d^{4}p}{(2\pi)^{4}}\widetilde{\tau}_{(0)}^{++}(p).
\label{delta_m}
\end{align}
The fact that all thermal propagators contain the same on-shell $\delta(p^{2}-m^{2})$ component leads to problems in Eq.~\eqref{1-loop++}, since the products of $\delta(p^{2}-m^{2})$ with itself as well as with the distribution $\frac{1}{p^{2}-m^{2}+i \epsilon}$ are mathematically ill-defined. However, in the difference these problematic products cancel, giving
\begin{align}
\widetilde{\tau}_{(0)}^{+-}(p)\widetilde{\tau}_{(0)}^{-+}(p)-\widetilde{\tau}_{(0)}^{++}(p)^{2} = \frac{1}{(p^{2}-m^{2}+i \epsilon)^{2}} +  \frac{2\pi i}{e^{\beta |p_{0}|}-1}\delta'(p^{2}-m^{2}),
\label{free_2}
\end{align}
and hence the one-loop result can be written in the form
\begin{align}
\widetilde{\tau}_{(1)}^{++}(p) = \widetilde{\tau}_{(0)}^{++}(p) + \Delta m_{+}^{2} \frac{\, \partial \widetilde{\tau}_{(0)}^{++} }{\partial m^{2}}(p).
\label{1-loop++:2}
\end{align}
The spectral function $\rho(p_{0},\vec{p})$ is defined
\begin{align}
\rho(p_{0},\vec{p}) = \int \! d^{4}x \ e^{ip \cdot x} \langle \Omega_{\beta}| [\phi(x),\phi(0)]|\Omega_{\beta}\rangle,
\end{align}
which implies the following general relation to the time-ordered propagator $\widetilde{\tau}^{++}(p)$ 
\begin{align}
\widetilde{\tau}^{++}(p) = i\int_{-\infty}^{\infty} \frac{dq_{0}}{2\pi} \frac{\rho(q_{0},\vec{p})}{|p_{0}| -q_{0} +i\epsilon} +  \frac{\rho(|p_{0}|,\vec{p})}{e^{\beta |p_{0}|}-1}.
\label{prop:rho}
\end{align}
Taking the real part of $\widetilde{\tau}^{++}(p)$, and using the known anti-symmetry of the spectral function in $p_{0}$, it follows from Eq.~\eqref{prop:rho} that
\begin{align}
\rho(p_{0},\vec{p}) = \epsilon(p_{0})\frac{\text{Re} \, \widetilde{\tau}^{++}(p)}{\frac{1}{2}\coth\left(\frac{\beta |p_{0}|}{2} \right)},
\label{rho:prop}
\end{align} 
where $\epsilon(p_{0})$ is the sign function. Using this relation, together with Eqs.~\eqref{1-loop++} and~\eqref{free_2}, the one-loop spectral function has the explicit form
\begin{align}
\rho_{(1)}(p_{0},\vec{p}) = 2\pi \epsilon(p_{0})\left[\delta(p^{2}-m^{2}) - \Delta m_{+}^{2} \delta'(p^{2}-m^{2}) \right].
\label{rho:1-loop}
\end{align}  
Equations~\eqref{1-loop++:2} and~\eqref{rho:1-loop} reflect the well-known result that at one-loop order the theory undergoes a thermal mass-shift $m^{2} \rightarrow m^{2} + \Delta m_{+}^{2}$. \\

\noindent
Although the total perturbative result is meaningful, it is problematic that each Feynman diagram contribution beyond leading order is guaranteed to contain ill-defined terms due to the presence of on-shell delta components in the basic field propagators. The standard strategy to avoid this issue is to start with a regularised form for the delta term in the propagators, for example choosing~\cite{Niemi:1983nf,Bellac:2011kqa,Das:1997gg}
\begin{align}
\delta_{\varepsilon}(x) = \frac{1}{\pi}\frac{\varepsilon}{x^{2}+\varepsilon^{2}}.
\label{epsilon_reg}
\end{align}
The problematic terms in the perturbative expansion are then well defined, and one can take the limit $\varepsilon \rightarrow 0^{+}$ at the end of the calculation. However, the fact that this regularisation needs to be imposed by hand is somewhat artificial, since it is only introduced to ensure the mathematical consistency of the computational procedure, and not for physical reasons. Even though the propagators can be arranged in a matrix form in which these problematic terms are manifestly absent~\cite{Bellac:2011kqa}, this does not alter the fact that they are present in the Wick contractions of each term in the series expansion of Eq.~\eqref{GML_Tv1:1}.

\item {\bf{Infrared divergences:}}
Another important issue in the standard approach to finite-temperature perturbation theory is the appearance of infrared divergent diagrams for massless theories. In $\phi^{4}$ theory the simplest such example is the cactus diagram\footnote{The cactus diagram is the middle diagram in Fig.~\ref{self_energy_diag}.}, which appears at two-loop order in the perturbative expansion of the propagator $\widetilde{\tau}(p)$. This diagram involves the component
\begin{align}
\int \frac{d^{4}p}{(2\pi)^{4}}\left(\frac{i}{(p^{2}-m^{2}+i \epsilon)^{2}} -  \frac{2\pi}{e^{\beta |p_{0}|}-1}\delta'(p^{2}-m^{2})  \right),
\label{cactus}
\end{align} 
which is finite after renormalisation when $m>0$, but diverges for $m=0$. The standard approach for dealing with these divergences is to sum these classes of diagrams to all loop orders, which is expected to result in finite contributions to the full perturbative series~\cite{Kapusta:2006pm,Bellac:2011kqa,Laine:2016hma}. Other approaches include the use of effective field theory~\cite{Braaten:1994na}, as well as various reorganisations of the perturbation expansion, such as optimised infinite resummations~\cite{Braaten:1989mz, Karsch:1997gj,Chiku:1998kd,Andersen:2000yj,Andersen:2008bz}, and variational techniques like the two-particle irreducible (2PI) formalism~\cite{Blaizot:2000fc,vanHees:2001ik,VanHees:2001pf,vanHees:2002bv}. However, although these approaches yield seemingly finite results, it has been demonstrated explicitly for massless scalar QFTs that the perturbative procedure explicitly breaks down at some fixed loop order because branch point singularities prohibit the computation of corrections to the propagator pole~\cite{Weldon:1998bj,Weldon:1998xr,Weldon:2001vt}. This breakdown is known to persist for the simplest resummation schemes~\cite{Weldon:2001vt}, and it remains unknown  whether any of the other proposed approaches can circumvent this problem. What remains clear is that the standard approach to perturbation theory based on the expansion of the thermal GML relation in Eq.~\eqref{GML_Tv1:1} has fundamental infrared problems, which for massless theories cannot be resolved by standard renormalisation arguments. 

\end{enumerate}

\noindent
Related to these more general problems, it is also the case that finite-temperature perturbative series in the standard approach show a drastically worse convergence behaviour than in vacuum, in spite of various refined technical improvements. This appears to be a generic feature which is independent of the dynamics, and hence similar for scalar theories as well as non-abelian gauge theories\footnote{For an early review see Ref.~\cite{Andersen:2004fp}.}. Recently, the QCD equation of state has been calculated on the lattice at temperatures up to the electroweak scale. Without fitting a $g^{6}$-coefficient to the lattice data, the analytic series are not consistent with the lattice data even at $T\sim 100$ GeV~\cite{Bresciani:2025vxw}, despite the fact that vacuum perturbation theory works perfectly at such high scales. Similarly, in scalar $\phi^{4}$ theory thermal perturbation theory is observed to increasingly worsen with temperature for coupling values where vacuum perturbation theory is quantitatively consistent~\cite{Lowdon:2024atn}.

\subsection{The imaginary-time formalism} 

Due to the relative complexity of the real-time formalism, perturbative calculations are often performed in imaginary time. For this purpose one makes use of the propagator $\widetilde{G}(k_{0},\vec{p})$ ($k_{0} \in \mathbb{C}$), which represents the analytic continuation of the real-time retarded and advanced propagators\footnote{In particular, the retarded and advanced propagators are recovered from $\widetilde{G}(k_{0},\vec{p})$ in the limit $k_{0}\rightarrow p_{0} \pm i\epsilon$~\cite{Lowdon:2022keu}.}. In the imaginary-time formalism the basic two-point correlation function appearing in the diagrams is the free-field Matsubara propagator 
\begin{align}
\widetilde{G}_{(0)}(k_{0}=i\omega_{N},\vec{p}) = \frac{1}{\omega_{N}^{2}+|\vec{p}|^{2} +m^{2}},
\label{0:loop:imag}
\end{align}
with $\omega_{N}=\tfrac{2\pi N}{\beta}$ ($N \in \mathbb{Z}$) the discrete Matsubara frequencies. In this case there are no on-shell singularities, and the complexity introduced by doubling the number of fields is replaced by the requirement to perform Matsubara sums instead of continuous energy integrals. In order to recover real-time results in the end one must perform an analytic continuation of the Matsubara propagator $i\omega_{N} \rightarrow p_{0} \pm i\epsilon$, from which one can determine the real-time retarded and advanced propagators, respectively. The simplest example is the one-loop calculation of the Matsubara propagator $\widetilde{G}_{(1)}(i\omega_{N},\vec{p})$, which has the explicit form
\begin{align}
\widetilde{G}_{(1)}(i\omega_{N},\vec{p}) = \widetilde{G}_{(0)}(i\omega_{N},\vec{p}) - \Delta m^{2} \widetilde{G}_{(0)}(i\omega_{N},\vec{p})^{2}.
\label{1-loop:imag}
\end{align}
In contrast to Eq.~\eqref{delta_m}, the computation of the bubble contribution $\Delta m^{2}$ requires one to perform a discrete Matsubara sum over $\widetilde{G}_{(0)}(i\omega_{N},\vec{p})$ instead of a continuous energy integral   
\begin{align}
\Delta m^{2} = \frac{\lambda}{2} \, T \! \sum_{N=-\infty}^{\infty} \int \frac{d^{3}\vec{p}}{(2\pi)^{3}} \widetilde{G}_{(0)}(i\omega_{N},\vec{p}).
\label{deltaM_imag}
\end{align}
Nevertheless, one finds that the real-time $\Delta m_{+}^{2}$ and imaginary-time $\Delta m^{2}$ bubble contributions coincide. The one-loop spectral function $\rho_{(1)}(p_{0},\vec{p})$ is computed as the difference between the one-loop retarded $\widetilde{r}_{(1)}(p)$ and advanced $\widetilde{a}_{(1)}(p)$ propagators
\begin{align}
\rho_{(1)}(p_{0},\vec{p}) &= -i\left[\widetilde{r}_{(1)}(p)-\widetilde{a}_{(1)}(p) \right] \nonumber \\
&= -i\lim_{\epsilon\rightarrow 0^{+}}\left[\widetilde{G}_{(1)}(p_{0}+i\epsilon,\vec{p})-\widetilde{G}_{(1)}(p_{0}-i\epsilon,\vec{p}) \right] \nonumber \\
&= \lim_{\epsilon\rightarrow 0^{+}}\left[-i\left(\widetilde{G}_{(0)}(p_{0}+i\epsilon,\vec{p})-\widetilde{G}_{(0)}(p_{0}-i\epsilon,\vec{p})\right)  \right. \nonumber \\
& \quad\quad\quad\quad\quad\quad\quad\quad\quad\quad\quad \left. + i \Delta m^{2}\left(\widetilde{G}_{(0)}(p_{0}+i\epsilon,\vec{p})^{2}-\widetilde{G}_{(0)}(p_{0}-i\epsilon,\vec{p})^{2}\right) \right]  \nonumber \\
&= 2\pi \epsilon(p_{0})\left[\delta(p^{2}-m^{2}) - \Delta m^{2} \delta'(p^{2}-m^{2}) \right].
\label{rho:1-loop:Im}
\end{align}  
In the final line we have made use of the fact that any power of the propagator $\widetilde{G}_{(0)}(k_{0},\vec{p})^{n}$ can be written in terms of derivatives of $\widetilde{G}_{(0)}(k_{0},\vec{p})$, and these derivatives commute with the limit $\lim_{\epsilon\rightarrow 0^{+}}$, which implies 
\begin{align}
\lim_{\epsilon\rightarrow 0^{+}}\widetilde{G}_{(0)}(p_{0} \pm i\epsilon,\vec{p})^{n} &= \lim_{\epsilon\rightarrow 0^{+}} (-1)^{n}\frac{(-1)^{n-1}}{(n-1)!}\frac{d^{n-1}}{d(p^{2}-m^{2})^{n-1}}
\frac{1}{(p^{2}-m^{2} \pm i\epsilon p_{0})} \nonumber \\
&= (-1)^{n}\frac{(-1)^{n-1}}{(n-1)!}\frac{d^{n-1}}{d(p^{2}-m^{2})^{n-1}} \lim_{\epsilon\rightarrow 0^{+}} \frac{1}{p^{2}-m^{2} \pm i\epsilon p_{0}} \nonumber \\
&= \text{p.v.}\frac{(-1)^{n}}{(p^{2}-m^{2})^{n}} \pm  \frac{1}{(n-1)!}i\pi \epsilon(p_{0})\delta^{(n-1)}(p^{2}-m^{2}),
\label{distr_power}
\end{align}
where we have used the distributional identity
\begin{align}
\lim_{\epsilon\rightarrow 0^{+}} \frac{1}{p^{2}-m^{2} \pm i\epsilon p_{0}} = \text{p.v.}\frac{1}{p^{2}-m^{2}} \mp i\pi \epsilon(p_{0})\delta(p^{2}-m^{2}),
\end{align}
and the notation $\delta^{(n)}(x) := \left(\frac{d}{dx}\right)^{n}\!\delta(x)$. The agreement between Eqs.~\eqref{rho:1-loop} and~\eqref{rho:1-loop:Im} is necessary to ensure that the real-time and imaginary-time formalisms give consistent perturbative results.

\subsection{Connection between the real and imaginary-time formalisms}
\label{diff:ReIm}

To understand why ill-defined on-shell terms appear in the real-time and not the imaginary-time formalism, one must consider how the basic field propagators are connected. From their definitions, the thermal time-ordered propagator $\widetilde{\tau}^{++}(p)$ and analytic propagator $\widetilde{G}(k_{0},\vec{p})$ are related in the following general manner
\begin{align}
\widetilde{\tau}^{++}(p) =-i\left(1+\frac{1}{e^{\beta |p_{0}|}-1} \right)\widetilde{G}(|p_{0}|+i0^{+},\vec{p})  + \frac{i}{e^{\beta |p_{0}|}-1}\widetilde{G}(|p_{0}|-i0^{+},\vec{p}),
\label{G_rel}
\end{align}
where $\widetilde{G}(|p_{0}|\pm i0^{+}, \vec{p}):=\lim_{\epsilon\rightarrow 0^{+}}\widetilde{G}(|p_{0}|\pm i\epsilon,\vec{p})$. Perturbative real-time calculations necessarily require one to take products of the basic field propagators. In the computation of $\widetilde{\tau}_{(1)}^{++}(p)$ in Eq.~\eqref{1-loop++} this involves the term $-\widetilde{\tau}_{(0)}^{++}(p)^{2}$, which due to Eq.~\eqref{G_rel} can be written 
\begin{align}
-\widetilde{\tau}_{(0)}^{++}(p)^{2} &= \widetilde{G}_{(0)}(|p_{0}|+i0^{+},\vec{p})^{2} + \frac{1}{e^{\beta |p_{0}|}-1}\left(\widetilde{G}_{(0)}(|p_{0}|+i0^{+},\vec{p})^{2}-\widetilde{G}_{(0)}(|p_{0}|-i0^{+},\vec{p})^{2} \right)   \nonumber \\
&  \quad\quad + \left(\frac{1}{e^{\beta |p_{0}|}-1} + \frac{1}{(e^{\beta |p_{0}|}-1)^{2}}\right)\!\left(\widetilde{G}_{(0)}(|p_{0}|+i0^{+},\vec{p})-\widetilde{G}_{(0)}(|p_{0}|-i0^{+},\vec{p}) \right)^{2}\!.
\label{t++} 
\end{align}  
Similarly, one can show that the other perturbative contribution $\widetilde{\tau}_{(0)}^{+-}(p)\widetilde{\tau}_{(0)}^{-+}(p)$ is given by
\begin{align}
\widetilde{\tau}_{(0)}^{+-}(p)\widetilde{\tau}_{(0)}^{-+}(p) &= - \left(\frac{1}{e^{\beta |p_{0}|}-1} + \frac{1}{(e^{\beta |p_{0}|}-1)^{2}}\right) \!\left(\widetilde{G}_{(0)}(|p_{0}|+i0^{+},\vec{p})-\widetilde{G}_{(0)}(|p_{0}|-i0^{+},\vec{p}) \right)^{2} \!.   
\label{t+-t-+}
\end{align} 
Comparing Eqs.~\eqref{t++} and~\eqref{t+-t-+} with Eq.~\eqref{1-loop:imag} one can immediately see that the extraction of real-time information from the imaginary-time perturbative expansion requires one to take the limit of products of the analytic propagator, which using Eq.~\eqref{distr_power} is always well defined. However, in the real-time formalism, for each Feynman diagram one has to instead consider the product of the propagator limits $\widetilde{G}_{(0)}(|p_{0}|\pm i0^{+},\vec{p})$. Whilst any pure power $\widetilde{G}_{(0)}(|p_{0}|\pm i0^{+},\vec{p})^{n}$ is well defined, and can be shown to coincide with Eq.~\eqref{distr_power}, the product of \textit{mixed} terms 
\begin{align}
\widetilde{G}_{(0)}(|p_{0}|+ i0^{+},\vec{p})\widetilde{G}_{(0)}(|p_{0}|- i0^{+},\vec{p})= \frac{1}{p^{2}-m^{2}+i0^{+}}\frac{1}{p^{2}-m^{2}-i0^{+}}
\label{pinch}
\end{align}
does not exist as a distribution~\cite{Brouder:2014hta}. These so-called pinch singularities appear in both the separate diagram contributions in Eq.~\eqref{t++} and~\eqref{t+-t-+}, but cancel when they are combined, resulting in the expression in Eq.~\eqref{free_2}. This cancellation of pinch singularities has been proven to hold true at each order of perturbation theory~\cite{Landsman:1986uw}. In the zero-temperature theory these problematic terms never arise since for $T\rightarrow 0$ the basic field propagator in Eq.~\eqref{G_rel} reduces to 
\begin{align}
\widetilde{\tau}_{(0)}^{++}(p) \rightarrow -i\widetilde{G}_{(0)}(|p_{0}|+i0^{+},\vec{p}) = \frac{i}{p^{2}-m^{2}+i0^{+}},
\end{align} 
and hence potential mixed terms are absent.

\subsection{Origin of the problems and previous propositions}
\label{origin}

It has been argued in the literature that the problems outlined in Sec.~\ref{probs} are actually a symptom of fundamental non-perturbative constraints that apply to finite-temperature QFTs. One crucial such constraint arises from the \textit{Narnhofer-Requardt-Thirring} (NRT) theorem~\cite{Narnhofer:1983hp}, which implies that any non-trivial scattering matrix $S$ at finite temperature cannot be constructed from scattering states which possess a dispersion relation $p_{0}=E(\vec{p})$, with $E(\vec{p})$ a real function. In particular, this means that neither free fields, nor quasi-particle-like propagators with real poles, can form the basis of a consistent finite-temperature perturbative expansion~\cite{Landsman:1988ta}. From a physical perspective, this constraint is due to the fact that the temperature effects of a thermal medium in equilibrium are everywhere present at all times. These thermal effects therefore need to be taken into account in the definition of \textit{every} thermal state, including scattering states at asymptotically large times. \\   

\noindent
At $T=0$ it is rigorously understood~\cite{Haag:1992hx} how the large-time scattering states in scalar QFTs are constructed from free fields $\phi_{0}(x)$, which appear in the $T=0$ GML relation in Eq.~\eqref{GML_0}. However, due to the NRT theorem this \textit{cannot} be the case when $T>0$, otherwise the full scattering matrix would be trivial, i.e. $S=1$. Therefore, the standard choice to define the thermal GML relation with $T=0$ scattering fields $\phi_{0}(x)$, as in Eq.~\eqref{GML_Tv1:1}, introduces a fundamental inconsistency into the perturbative formalism. In fact, both the appearance of pinch singularities in Feynman diagrams and the existence of infrared divergences can be traced back to the assumption that the thermal asymptotic scattering states are on shell with $p^{2}=m^{2}$. Interestingly, this infrared problem is analogous to what happens in Quantum Electrodynamics (QED), where infrared divergences occur in the S-matrix when one chooses naive on-shell scattering states for the electrons. By correctly taking into account the presence of soft photons in the definition of these states, which must exist by virtue of the electron's charge, one can demonstrate that the corresponding S-matrix becomes finite~\cite{Strocchi:2013awa}.  \\

\noindent   
It is clear that a consistent finite-temperature perturbative framework must be based on a thermal generalisation of the GML relation which avoids the non-perturbative constraints imposed by the NRT theorem, and therefore requires the construction of appropriate finite-temperature scattering states. A significant attempt in this direction was made in Ref.~\cite{Landsman:1988ta}, where the thermal scattering states were parametrised in terms of resonance-like modes with poles in the (retarded) propagator of the form
\begin{align}
k_{0}=E(\vec{p})-i\Gamma(\vec{p}), 
\label{C_disp} 
\end{align}
with $E(\vec{p})$ and $\Gamma(\vec{p})$ some real-valued functions. By not having real dispersion relations these states avoid the constraints imposed by the NRT theorem, and open the possibility for defining a consistent finite-temperature perturbative framework~\cite{Landsman:1988ta}. Similar ideas were also put forward in Ref.~\cite{Weldon:1998bj}. Although this approach has several desirable characteristics, there still remain some open issues. Perhaps the most pressing problem is that whilst the functions $E(\vec{p})$ and $\Gamma(\vec{p})$ in Eq.~\eqref{C_disp} must depend on the dynamics of the full theory, it is not clear what determines their parametric structure. Any perturbative expansion therefore requires a choice of ansatz for the asymptotic state propagators instead of this being fixed by the fundamental theory. \\

\noindent
Another problem with choosing propagators that have the singularity structure in Eq.~\eqref{C_disp} is that their form coincides with those of resonance-like states. In particular, when $\Gamma$ is a constant and $E(\vec{p})=\sqrt{|\vec{p}|^{2}+m^{2}}$ the propagator reduces to that of a relativistic Breit-Wigner, whose spectral function has the following explicit form
\begin{align}
\rho_{\text{BW}}(p_{0},\vec{p}) = \frac{4  p_{0} \Gamma}{(p_{0}^{2}-|\vec{p}|^{2}-m^{2}-\Gamma^{2})^{2} + 4 p_{0}^{2}\, \Gamma^{2}}.
\label{rho_BW:p}
\end{align}  
$\rho_{\text{BW}}(p_{0},\vec{p})$ has a broadened peak-like structure, which is qualitatively what one would expect in the presence of interactions with the thermal medium. However, in position space: $\rho_{\text{BW}}(x_{0},\vec{x}) = e^{-\Gamma |x_{0}|}\Delta(x;m)$, where $\Delta(x;m)$ is the commutator function for a free particle of mass $m$, and hence along any direction $\vec{x}=\vec{v}x_{0}$ these states decay rapidly in time irrespective of their velocity $\vec{v}$ relative to the medium. This intrinsically unstable behaviour is not consistent with the physical picture of stable particle-like states moving in a medium, where the thermal effects experienced by these states are frame-dependent and must ultimately vanish on average in the limit $\vec{v}\rightarrow 0$~\cite{Bros:1996mw,Bros:2003zs}.

\section{Thermoparticle perturbation theory}
\label{TP_PT}

\subsection{Thermoparticles}
\label{TP}

The physical and structural deficiencies of the finite-temperature approaches based on states with complex propagator poles of the form in Eq.~\eqref{C_disp} prompted a series of studies~\cite{Bros:1992ey,Buchholz:1993kp,Bros:1995he,Bros:1996mw,Bros:2001zs,Bros:2003zs} which establish the basic non-perturbative properties satisfied by correlation functions in finite-temperature QFTs, and put fundamental constraints on the spectral characteristics of thermal equilibrium systems. A highly significant result in this regard was the derivation of the spectral representation satisfied by the thermal two-point correlation functions of scalar field operators. In particular, the authors proved~\cite{Bros:1996mw} that the spectral function $\rho(p_{0},\vec{p})$ has the general representation\footnote{For a more recent discussion of this representation see Ref.~\cite{Nair:2025jgl}.} 
\begin{align}
\rho(p_{0},\vec{p}) = \int_{0}^{\infty} \! ds \int \! \frac{d^{3}\vec{u}}{(2\pi)^{2}} \ \epsilon(p_{0}) \, \delta\!\left(p_{0}^{2} - (\vec{p}-\vec{u})^{2} - s \right)\widetilde{D}_{\beta}(\vec{u},s).    
\label{rho_rep}
\end{align}  
An immediate implication of Eq.~\eqref{rho_rep} is that the temperature and dynamical dependence of the spectral function is entirely encoded in the thermal spectral density $\widetilde{D}_{\beta}(\vec{u},s)$. Determining the structure of $\widetilde{D}_{\beta}(\vec{u},s)$ is therefore essential for establishing the fundamental characteristics of thermal excitations in the theory. In Ref.~\cite{Bros:1992ey} it was argued that if the vacuum theory contains a stable particle state of mass $m$, then the thermal spectral density $\widetilde{D}_{\beta}(\vec{u},s)$ must contain a distinguished component of the form 
\begin{align}
\widetilde{D}_{m,\beta}(\vec{u})\, \delta(s-m^{2}).
\label{TP_damping}
\end{align}
In the $T\rightarrow 0$ limit the restoration of Lorentz symmetry implies: $\widetilde{D}_{m,\beta}(\vec{u}) \rightarrow (2\pi)^{3}\delta^{3}(\vec{u})$, and after substitution into Eq.~\eqref{rho_rep} one recovers the spectral function for a massive particle state in vacuum, 
\begin{align}
\rho(p_{0},\vec{p})= 2\pi \epsilon(p_{0})\delta(p^{2}-m^{2}).
\label{rho0} 
\end{align}
Equation~\eqref{TP_damping} therefore represents the finite-temperature generalisation of stable particle states. In order to draw a clear distinction with other potential thermal excitations, such as collective quasi-particle modes which vanish when $T\rightarrow 0$, these components were subsequently referred to as \textit{thermoparticles}~\cite{Buchholz:1993kp}. On account of the representation in Eq.~\eqref{rho_rep}, when $\widetilde{D}_{m,\beta}(\vec{u})$ has a non-trivial structure this implies that the vacuum $p^{2}=m^{2}$ peak in Eq.~\eqref{rho0} becomes broadened. This broadening in momentum-space represents the effect caused by the collisional interactions with the medium, since any transfer of energy to the thermal background will cause particle states to go off their mass shell. Equivalently, in position space the temperature effects are controlled by a multiplicative factor $D_{m,\beta}(\vec{x})$, the inverse Fourier transform of $\widetilde{D}_{m,\beta}(\vec{u})$. Since $D_{m,\beta}(\vec{x})$ decreases in amplitude with increasing temperature this inhibits the propagation of the state, lowering its mean-free path, and hence $D_{m,\beta}(\vec{x})$ has the physical interpretation of a thermal damping factor. \\

\noindent
Due to their characteristic form in Eq.~\eqref{TP_damping} thermoparticles have a number of distinctive properties:

\begin{itemize}

\item Thermoparticle spectral functions $\rho_{\text{TP}}(p_{0},\vec{p})$ have an energy threshold at the corresponding vacuum particle mass $|p_{0}|=m$, and hence one can always write
\begin{align}
\rho_{\text{TP}}(p_{0},\vec{p}) = \theta(p_{0}^{2}-m^{2})\varrho(p_{0},\vec{p}),
\end{align}
where $\varrho(p_{0},\vec{p})$ is some distribution restricted to $|p_{0}|\geq m$. Physically, this implies that the thermal medium must be excited with an energy of at least $m$ in order to create a thermoparticle state with a non-vanishing probability. 

\item The thermoparticle two-point function $\mathcal{W}_{\text{TP}}(x_{0},\vec{x})$ has the factorised form
\begin{align}
\mathcal{W}_{\text{TP}}(x_{0},\vec{x}) = D_{m,\beta}(\vec{x}) \mathcal{W}_{(0)}(x_{0},\vec{x}),
\label{2pt_TP}
\end{align}
where $\mathcal{W}_{(0)}(x_{0},\vec{x})$ is the thermal two-point function for a non-interacting mass $m$ field\footnote{This thermal two-point function is defined: $ \mathcal{W}_{(0)}(x_{0},\vec{x}) = \int \frac{d^{4}p}{(2\pi)^{4}}e^{-ip \cdot x} \, 2\pi\epsilon(\omega) \, \delta\!\left(\omega^{2} - |\vec{p}|^{2} - m^{2} \right)(1-e^{-\beta\omega})^{-1}$.}. Along the direction $\vec{x}=\vec{v}x_{0}$ Eq.~\eqref{2pt_TP} implies that thermoparticles have the large-time asymptotic behaviour 
\begin{align}
\mathcal{W}_{\text{TP}}(x_{0},\vec{x}) \sim D_{m,\beta}(\vec{v}x_{0})|x_{0}|^{-\frac{3}{2}}, \quad |x_{0}| \rightarrow \infty.
\end{align}
Any damping these states experience therefore depends on their velocity relative to the medium~\cite{Bros:2003zs}, which is not the case for resonance-like states, as discussed in Sec.~\ref{origin}. In particular, since the damping factor satisfies the temperature-independent normalisation condition $D_{m,\beta}(0)=1$~\cite{Bros:2001zs}, it follows that thermoparticles at rest ($\vec{v}=0$) behave on average like they are not subject to collisions with the medium, as one would physically expect~\cite{Bros:2003zs}.

\item Thermoparticles \textit{dominate} the large-time behaviour of thermal correlation functions, and in particular have the same asymptotic decay as vacuum particle correlators~\cite{Bros:2001zs}.

\item Thermoparticle states avoid the constraints imposed by the NRT theorem, since the presence of a non-trivial damping factor $D_{m,\beta}(\vec{x})$ prevents thermoparticles from having a purely-real dispersion relation. 

\end{itemize}

\noindent
Given these characteristic properties, thermoparticles are a natural candidate for describing finite-temperature scattering states. In Ref.~\cite{Bros:2001zs} the authors further proposed a consistency condition\footnote{See Ref.~\cite{Bros:2001zs} for details regarding the technical implementation of this condition.} for the asymptotic fields $\phi_{\text{TP}}(x)$ which generate these states: 
\begin{align}
\phi_{\text{TP}}(x) \ \textit{must satisfy the equation of motion of the theory for} \  |x_{0}|\rightarrow \infty.
\label{asymptot_cond}
\end{align}
This constraint implies that even at large times the thermal medium encodes information about the dynamics of the theory, and for thermoparticle states it turns out that this \textit{uniquely} fixes the form of their corresponding damping factor $D_{m,\beta}(\vec{x})$~\cite{Bros:2001zs}. Since the asymptotic fields themselves determine the structure of the basic propagators which appear in the perturbative expansion of the GML relation, the approach of Ref.~\cite{Bros:2001zs} defines a procedure by which the form of these propagators can be self-consistently derived from the dynamical equations of the QFT, in contrast to the approaches outlined in Sec.~\ref{origin}. Ultimately, if thermoparticle states are indeed the correct basis for performing finite-temperature perturbative expansions, different theories will necessarily require distinct propagators. This is physically reasonable, since finite-temperature phenomena are strongly dependent on the dynamics of the underlying medium. \\

\noindent
As well as having compelling theoretical properties, there is now also extensive evidence for the existence of thermoparticle states in a variety of rather different finite-temperature QFTs, including real~\cite{Lowdon:2024atn} and complex~\cite{Lowdon:2025fyb,Lowdon:2025ait} scalar theories with and without spontaneous symmetry breaking, as well as quantum chromodynamics (QCD)~\cite{Lowdon:2022xcl,Bala:2023iqu}. In each of these studies, it was demonstrated using non-perturbative lattice QFT simulations that the correlation functions are consistent with the presence of these excitations, and that for temperatures of the order of the vacuum particle mass they play a dominant role in determining the low-energy spectral properties of the theories. Taken together, these findings strongly indicate that the finite-temperature generalisation of the GML relation should be based on thermoparticle states.

\subsection{Finite-temperature generalisation of the Gell-Mann-Low relation}
\label{GML_rel_TP}

Another more subtle issue raised in Ref.~\cite{Landsman:1988ta} is the use of the thermal trace $\text{Tr}(e^{-\beta H} \cdots)$ in Eq.~\eqref{GML_Tv1:1} to define the thermal expectation values of the fields. Rigorously speaking, the trace of the operator $e^{-\beta H}$ only exists for finite volumes, since the Hamiltonian operator $H$ in this case has a positive discrete spectrum~\cite{Landsman:1988ta}. For infinite-volume QFTs the notion of thermal equilibrium therefore needs to be implemented in a different manner. A resolution to this problem is to demand that the thermal correlation functions satisfy the Kubo-Martin-Schwinger (KMS) condition~\cite{Haag:1967sg} 
\begin{align}
&\langle \Omega_{\beta}|\phi(x_{1})\cdots \phi(x_{k})\phi(x_{k+1})\cdots \phi(x_{n})|\Omega_{\beta}\rangle  \nonumber \\
& \quad\quad\quad\quad\quad\quad\quad\quad\quad\quad = \langle \Omega_{\beta}|\phi(x_{k+1})\cdots \phi(x_{n}) \phi(x_{1}+i(\beta,\vec{0}))\cdots \phi(x_{k}+i(\beta,\vec{0}) )|\Omega_{\beta}\rangle.
\label{KMS}
\end{align}
Equation~\eqref{KMS} holds true in the infinite-volume limit, and can be shown to be equivalent to the canonical thermal trace definition for finite-volume systems~\cite{Haag:1967sg}. In order to avoid these ambiguities one should therefore replace the thermal trace in Eq.~\eqref{GML_Tv1:1} with an expectation value with respect to a thermal ground state whose correlation functions satisfy the KMS condition. \\

\noindent 
If thermoparticles do indeed provide the correct description of thermal scattering states, as indicated in Ref.~\cite{Bros:2001zs}, this suggests that the GML relation should have the following finite-temperature generalisation
\begin{align}
\langle \Omega_{\beta}|T\!\left\{ \phi(x_{1})\cdots \phi(x_{n}) \right\}|\Omega_{\beta}\rangle = \frac{\langle \Omega_{\beta}^{0}|T_{C}\!\left\{ \phi_{\text{TP}}(x_{1})\cdots \phi_{\text{TP}}(x_{n}) e^{i \int_{C} \, \mathcal{L}_{I}\left[\phi_{\text{TP}}\right]} \right\} | \Omega_{\beta}^{0}\rangle }{ \langle \Omega_{\beta}^{0}|T_{C}\!\left\{ e^{i \int_{C} \, \mathcal{L}_{I}\left[\phi_{\text{TP}}\right]} \right\} | \Omega_{\beta}^{0}\rangle}, 
\label{GML_TP}
\end{align}
where $\phi_{\text{TP}}(x)$ is the asymptotic field associated with these states, and $|\Omega_{\beta}^{0}\rangle$ is the KMS ground state of the scattering spectrum generated by these fields. This proposed generalisation is analogous to the one derived in Ref.~\cite{Landsman:1988ta}, except that the choice of asymptotic scattering fields $\phi_{\text{TP}}(x)$ have very different properties to those used in Ref.~\cite{Landsman:1988ta}. A fundamental difference to the vacuum GML relation in Eq.~\eqref{GML_0} is that $\phi_{\text{TP}}(x)$ is not a free field, and hence the correlation functions appearing in the perturbative expansion of Eq.~\eqref{GML_TP} encode information about the dynamics of the theory. Nevertheless, it turns out that $\phi_{\text{TP}}(x)$ define \textit{quasi-free} states, and hence their correlation functions can always be decomposed into sums of products of propagators, just like at zero temperature~\cite{Bros:2001zs}. The resulting terms that appear in the perturbative expansion of Eq.~\eqref{GML_TP} can therefore be decomposed into components with exactly the same topology as in the standard finite-temperature approach, the only difference is that the propagators appearing in these expressions now have the general structure
\begin{align}
\langle \Omega_{\beta}^{0}|T\left\{ \phi_{\text{TP}}^{a}(x)\phi_{\text{TP}}^{b}(0)\right\} | \Omega_{\beta}^{0}\rangle = D_{m,\beta}(\vec{x}) \tau_{(0)}^{ab}(x), 
\label{TP_prop}
\end{align}
where $\tau_{(0)}^{ab}(x)$ have the momentum-space forms in Eqs.~\eqref{free_prop} and~\eqref{other_prop} with $a,b=+,-$.

\subsection{Resolving the inconsistencies of the standard approach}

Starting from the generalised GML relation Eq.~\eqref{GML_TP}, and restricting both fields to $C_{+}$, it follows from Eq.~\eqref{prop:rho} and the results in Sec.~\ref{TP} that the modified real-time propagator has the form   
\begin{align}
\widetilde{\tau}_{\text{TP}}^{++}(p) = \int \! \frac{d^{3}\vec{u}}{(2\pi)^{3}} \left[\frac{i}{p_{0}^{2} - (\vec{p}-\vec{u})^{2} - m^{2} +i\epsilon }  + \frac{2\pi \delta\!\left(p_{0}^{2}-(\vec{p}-\vec{u})^{2} - m^{2} \right)}{e^{\beta|p_{0}|}-1} \right] \! \widetilde{D}_{m,\beta}(\vec{u}). 
\label{prop:rho:TP}
\end{align} 
Comparing this expression with Eq.~\eqref{free_prop} one immediately sees a fundamental difference: due to the appearance of the damping factor the propagator no longer decomposes into temperature-independent and dependent components. The physical interpretation of the standard approach, that the system is comprised of separate quantum and heat bath sub-systems~\cite{Das:1997gg}, is therefore lost. This is a reflection of the fact that even at large times the thermal medium contains information about the dynamics of the system, and hence temperature effects never decouple within the propagator of the asymptotic states. Similarly, in the imaginary-time formalism the corresponding Matsubara propagator is given by~\cite{Lowdon:2022keu}
\begin{align}
\widetilde{G}_{\text{TP}}(i\omega_{N},\vec{p}) &= \int \! \frac{d^{3}\vec{u}}{(2\pi)^{3}}   \, \frac{1}{\omega_{N}^{2} + (\vec{p}-\vec{u})^{2} + m^{2} } \, \widetilde{D}_{m,\beta}(\vec{u}).
\label{Matsubara:TP}
\end{align} 
From Eqs.~\eqref{prop:rho:TP} and~\eqref{Matsubara:TP} it follows that the perturbative propagators of the standard approach in Eqs.~\eqref{free_prop} and~\eqref{0:loop:imag} are recovered for $\widetilde{D}_{m,\beta}(\vec{u}) \rightarrow (2\pi)^{3}\delta^{3}(\vec{u})$, which corresponds to the limit where the asymptotic fields are free from interactions and on shell, i.e. $\phi_{\text{TP}}(x) \rightarrow \phi_{0}(x)$. In the thermoparticle case the damping factor $\widetilde{D}_{m,\beta}(\vec{u})$ has a non-trivial $\vec{u}$ dependence, which upon integration results in a significantly modified singularity structure of both propagators. This property is highly consequential, and provides a resolution to two of the major problems of the standard approach outlined in Sec.~\ref{probs}:

\begin{enumerate}[leftmargin=*]

\item {\bf{The absence of ill-defined on-shell components:}} Due to the non-trivial behaviour of $\widetilde{D}_{m,\beta}(\vec{u})$, the would-be on-shell delta term in Eq.~\eqref{prop:rho:TP} is replaced by a contribution which has a broadened peak-like structure around the vacuum singularity $p^{2}=m^{2}$. The broadening guarantees that this propagator component can be consistently multiplied with itself, and hence the contributions appearing in the real-time diagrammatic expansion are mathematically well defined. In particular, the pinch singularity defined in Eq.~\eqref{pinch} is absent because the broadening effects ensure that the propagators in the product $\widetilde{G}_{\text{TP}}(|p_{0}|+ i0^{+},\vec{p})\widetilde{G}_{\text{TP}}(|p_{0}|- i0^{+},\vec{p})$ no longer have on-shell singularities. In contrast to the standard strategy, where on-shell delta terms are regularised by hand as in Eq.~\eqref{epsilon_reg}, in the present formalism the regularisation occurs through physical principles, namely that at large times the thermal medium still contains information about the dynamics of the system. Since these dynamics are only absent for the asymptotic states in the $T\rightarrow 0$ limit, this regularisation via thermal damping is always present, as opposed to the standard approach, where the artificial width is taken to zero at the end of the calculation. This suggests that the regularisation of pinch singularities in the standard perturbative approach may not merely be a technical inconvenience, but possibly interfere with the physics of the system.

\item {\bf{The absence of infrared divergences:}} In the standard perturbative approach infrared divergences are most easily understood in the imaginary-time formalism, since they are known to arise in terms that involve the zero Matsubara frequency modes. For example, in massless $\phi^{4}$ theory the cactus diagram component in Eq.~\eqref{cactus} can be written     
\begin{align}
T\sum_{N=-\infty}^{\infty} \int \frac{d^{3}p}{(2\pi)^{3}}\frac{1}{(\omega_{N}^{2}+|\vec{p}|^{2})^{2}}.
\label{cactus:imag}
\end{align}
Each individual term in the series is finite except $N=0$, which is infrared divergent. If one instead considers the same calculation with a massless thermoparticle propagator, it follows from the form of Eq.~\eqref{Matsubara:TP} that the zero-mode contribution can be written 
\begin{align}
&T \int \! \frac{d^{3}p}{(2\pi)^{3}}\int \! \frac{d^{3}\vec{u}}{(2\pi)^{3}}\int \! \frac{d^{3}\vec{v}}{(2\pi)^{3}}   \, \frac{\widetilde{D}_{0,\beta}(\vec{u})\widetilde{D}_{0,\beta}(\vec{v})}{(\vec{p}-\vec{u})^{2}(\vec{p}-\vec{v})^{2}}  \nonumber \\
&= T \int \! \frac{d^{3}p}{(2\pi)^{3}} \int_{0}^{\infty} \!d|\vec{u}| \int_{0}^{\infty} \! d|\vec{v}| \, \frac{|\vec{u}||\vec{v}|}{64\pi^{4}|\vec{p}|^{2}} \ln \!\left[\frac{(|\vec{p}|+|\vec{u}|)^{2}}{(|\vec{p}|-|\vec{u}|)^{2}}  \right]  \ln \!\left[\frac{(|\vec{p}|+|\vec{v}|)^{2}}{(|\vec{p}|-|\vec{v}|)^{2}}  \right] \widetilde{D}_{0,\beta}(\vec{u})\widetilde{D}_{0,\beta}(\vec{v}),
\label{cactus:imag:TP}
\end{align} 
where in the final line the angular integrations have been performed explicitly since the damping factors only depend on the absolute value of $\vec{u}$ and $\vec{v}$. Note that because the damping factors are regular at the point $\vec{u}=\vec{v}=0$, which is \textit{not} the case for free scattering fields $\phi_{0}(x)$, since $\widetilde{D}_{0,\beta}(\vec{u}) = (2\pi)^{3}\delta^{3}(\vec{u})$, one can expand the integrand around the point $\vec{p}=0$ to elucidate the infrared behaviour of the momentum integral,
\begin{align}
\frac{|\vec{u}||\vec{v}|}{64\pi^{4}|\vec{p}|^{2}} \ln \!\left[\frac{(|\vec{p}|+|\vec{u}|)^{2}}{(|\vec{p}|-|\vec{u}|)^{2}}  \right]  \ln \!\left[\frac{(|\vec{p}|+|\vec{v}|)^{2}}{(|\vec{p}|-|\vec{v}|)^{2}}  \right]  = \frac{1}{4\pi^{4}} + \mathcal{O}(|\vec{p}|^{2}).
\end{align} 
The $1/|\vec{p}|^{4}$ behaviour of the integrand which leads to the divergence in Eq.~\eqref{cactus:imag} for $N=0$ is therefore no longer present in the thermoparticle case. This screening of the infrared pole is brought about by physical principles rather than an artificial regularisation: the presence of the damping factor integration shifts the $\vec{p}=0$ pole is away from the real axis, resulting in a finite momentum integral. Whilst Eq.~\eqref{cactus:imag:TP} corresponds to one specific example, this screening of infrared poles is a generic feature which will apply to any diagram that involves zero mode contributions. 

\end{enumerate}

\subsection{Perturbative expansion and renormalisability}
\label{PT_ren}

A well-known characteristic of any QFT is the appearance of ultraviolet divergences in the perturbative expansion. A theory is said to be perturbatively renormalisable if these divergences can be regularised in a systematic manner order-by-order in the expansion. For zero-temperature QFTs, expanding the exponential in the GML relation in Eq.~\eqref{GML_0} gives rise to the perturbative series 
\begin{align}
\langle \Omega|T\!\left\{ \phi(x_{1})\cdots \phi(x_{n}) \right\}|\Omega\rangle = \sum_{N=0}^{\infty} \tau^{(N)}(x_{1}, \dots, x_{n}),
\label{PT_0}
\end{align}
where the order-$N$ contribution has the form
\begin{align}
\tau^{(N)}(x_{1}, \dots, x_{n}) = \tfrac{i^{N}}{N! \mathcal{Z}}\int \! d^{4}z_{1}\cdots d^{4}z_{N}  \langle \Omega_{0}|T\!\left\{ \phi_{0}(x_{1})\cdots \phi_{0}(x_{n}) \mathcal{L}_{I}[\phi_{0}(z_{1})]\cdots \mathcal{L}_{I}[\phi_{0}(z_{N})] \right\} | \Omega_{0} \rangle,   
\label{GML_0_2}
\end{align}
and $\mathcal{Z} = \langle \Omega_{0}|T\{ e^{i \int d^{4}z \, \mathcal{L}_{I}\left[\phi_{0}(z)\right]} \} | \Omega_{0}\rangle$ is expanded to the order in the coupling at which the series is truncated. From the structure of Eq.~\eqref{GML_0_2} it follows that ultraviolet divergences arise either from the definition of $\mathcal{L}_{I}[\phi_{0}(z_{i})]$, which involves an a priori ill-defined product of fields $\phi_{0}(z_{i})$, or when any two or more of the terms in the product $\mathcal{L}_{I}[\phi_{0}(z_{1})]\cdots \mathcal{L}_{I}[\phi_{0}(z_{N})]$ have coinciding spacetime arguments~\cite{Haag:1992hx}. Since $\mathcal{L}_{I}[\phi_{0}(z_{i})]$ involves the product of free fields, this can be given a well-defined meaning via Wick (i.e. normal) ordering. In the case of $\phi^{4}$ theory one therefore defines 
\begin{align}
\mathcal{L}_{I}[\phi_{0}(z_{i})] = -\frac{\lambda}{4!} \!:\!\phi_{0}^{4}\!:\!(z_{i}).
\end{align}
Once the definition of $\mathcal{L}_{I}$ is set, the question of renormalisability reduces to whether the product of these operators in the limit of coincident arguments is proportional to some other operator within a finite set, where the constant of proportionality formally diverges. In other words, whether any diagram only contains a finite class of sub-divergences. If so, then the theory can be renormalised by replacing at each perturbative order these divergent constants with finite coefficients, which must be fixed by some renormalisation condition. Proving the renormalisability of any given QFT is a highly non-trivial endeavour, particularly at finite temperature, where one must demonstrate that temperature effects do not lead to additional ultraviolet divergences~\cite{Landsman:1986uw}. In the remainder of this section we will outline why thermoparticle perturbation theory avoids such contributions, which suggests that the renormalisation of the $T=0$ theory should be sufficient to guarantee finite perturbative predictions for any $T>0$.  \\  

\noindent
The analogue of Eq.~\eqref{GML_0_2} in the thermoparticle perturbation theory case is
\begin{align}
\frac{i^{N}}{N! \mathcal{Z}}\int_{C_{1}} \! d^{4}z_{1} \cdots  \int_{C_{N}} \! d^{4}z_{N}  \ \langle \Omega_{\beta}^{0}|T_{C}\!\left\{ \phi_{\text{TP}}(x_{1})\cdots \phi_{\text{TP}}(x_{n}) \mathcal{L}_{I}[\phi_{\text{TP}}(z_{1})]\cdots \mathcal{L}_{I}[\phi_{\text{TP}}(z_{N})] \right\} | \Omega_{\beta}^{0}\rangle,
\label{GML_TP_N}
\end{align}
where the integration along the Schwinger-Keldish contours $C_{i}$ are defined in Eq.~\eqref{SK_int}. As outlined in Sec.~\ref{GML_rel_TP}, the thermoparticle fields encode information about the dynamics of the theory, and hence are no longer free fields. This means that unlike the $T=0$ case or the standard $T>0$ approach, the composite operators $\mathcal{L}_{I}[\phi_{\text{TP}}(z_{i})]$ cannot be defined via a Wick ordering. However, one \textit{can} define the product of thermoparticle fields via a point-splitting regularisation, which ensures that the field operators are temperature independent~\cite{Bros:2001zs}. In the case of two fields this is defined
\begin{align}
&\phi_{\text{TP}}^{2}(x) = \lim_{\varepsilon \rightarrow 0}\left[ \phi_{\text{TP}}(x+ \varepsilon)\phi_{\text{TP}}(x- \varepsilon) - \langle \Omega | \phi_{\text{TP}}(x+ \varepsilon)\phi_{\text{TP}}(x- \varepsilon) | \Omega \rangle \cdot 1  \right], 
\end{align}    
and for higher powers it is defined similarly\footnote{More details regarding this procedure can be found in Ref.~\cite{Bros:2001zs}.}. In the case of $\phi^{4}$ theory this therefore allows one to make sense of $\mathcal{L}_{I}[\phi_{\text{TP}}(z_{i})]= -\frac{\lambda}{4!} \phi_{\text{TP}}^{4}(z_{i})$. As in the zero-temperature case, the other source of ultraviolet divergences is when the product of two or more $\mathcal{L}_{I}[\phi_{\text{TP}}(z_{i})]$ terms in Eq.~\eqref{GML_TP_N} have coinciding spacetime arguments. Since the correlation function integrand in Eq.~\eqref{GML_TP_N} can always be decomposed into sums of products of two-point propagators, these singular terms arise from \textit{powers} of the basic thermoparticle propagator. Due to the structure of the propagator in Eq.~\eqref{TP_prop}, along the forward real-time branch $C_{+}$ these terms can be written in the form    
\begin{align}
\langle \Omega_{\beta}^{0}|T\!\left\{ \phi_{\text{TP}}^{+}(z_{i})\phi_{\text{TP}}^{+}(z_{j})\right\} | \Omega_{\beta}^{0}\rangle^{n}= D_{m,\beta}(\vec{z}_{i}-\vec{z}_{j})^{n} \tau_{(0)}^{++}(z_{i}-z_{j})^{n},
\label{PT_N}
\end{align} 
where $\tau_{(0)}^{++}(z)$ is the free-field propagator of the standard real-time approach. Due to the temperature-independent normalisation condition $D_{m,\beta}(0)=1$ it immediately follows that 
\begin{align}
\lim_{z_{i}\rightarrow z_{j}}\langle \Omega_{\beta}^{0}|T\!\left\{ \phi_{\text{TP}}^{+}(z_{i})\phi_{\text{TP}}^{+}(z_{j})\right\} | \Omega_{\beta}^{0}\rangle^{n}= \lim_{z_{i}\rightarrow z_{j}} \tau_{(0)}^{++}(z_{i}-z_{j})^{n}.
\label{PT_N_lim}
\end{align}    
Hence the ultraviolet singularities which arise from the coincident limit have exactly the same structure as in the standard approach. The presence of the damping factor therefore does not alter the ultraviolet behaviour of the correlation functions at any perturbative loop order, and so the renormalisation of divergences in the $T=0$ theory should be sufficient to guarantee the finiteness of all perturbative predictions when $T>0$.

\subsection{One-loop example} 
\label{one-loop}

To explore the differences between thermoparticle perturbation theory and the standard finite-temperature approach we consider the simplest example, the one-loop calculation of the scalar time-ordered propagator in $\phi^{4}$ theory. Since the topology of the diagrams in the perturbative expansion remains unchanged, the one-loop momentum space propagator has an analogous structure to Eq.~\eqref{1-loop++}
\begin{align}
\widetilde{\tau}_{(1), \text{TP}}^{++}(p) = \widetilde{\tau}_{\text{TP}}^{++}(p) + i \Delta m_{\text{TP}}^{2} \left(\widetilde{\tau}_{\text{TP}}^{+-}(p)\widetilde{\tau}_{\text{TP}}^{-+}(p)-\widetilde{\tau}_{\text{TP}}^{++}(p)^{2} \right),  
\label{GML_TP_1}
\end{align}
where the corresponding thermal bubble contribution $\Delta m_{\text{TP}}^{2}$ is now given by
\begin{align}
\Delta m_{\text{TP}}^{2} = \frac{\lambda}{2} \int \frac{d^{4}p}{(2\pi)^{4}}\left[\widetilde{\tau}_{\text{TP}}^{++}(p)-\widetilde{\tau}_{\text{TP}}^{++}(p;T=0) \right].
\label{delta_m_TP}
\end{align}
The subtraction of the $T=0$ propagator in Eq.~\eqref{delta_m_TP} arises from the point-splitting regularisation used to define $\mathcal{L}_{I}=-\frac{\lambda}{4!} \phi_{\text{TP}}^{4}$, and cancels the UV divergence. Although the thermoparticle propagator possesses a non-trivial damping factor dependence, this dependence drops out in $\Delta m_{\text{TP}}^{2}$ because the momentum integral of $\widetilde{\tau}_{\text{TP}}^{++}(p)$ coincides with the position space propagator at zero separation, which due to Eq.~\eqref{PT_N_lim} for $n=1$ coincides with the free-field result, and hence
\begin{align}
\Delta m_{\text{TP}}^{2} &= \frac{\lambda}{2} \int \frac{d^{4}p}{(2\pi)^{4}}\left[\widetilde{\tau}_{(0)}^{++}(p)-\widetilde{\tau}_{(0)}^{++}(p;T=0) \right] = \frac{\lambda}{2}\int \! \frac{d^{3}\vec{p}}{(2\pi)^{3}}\frac{1}{\sqrt{|\vec{p}|^{2}+m^{2}}}\frac{1}{e^{\beta\sqrt{|\vec{p}|^{2}+m^{2}}}-1},
\label{delta_m_TP_2}
\end{align}
which is finite. Thus, a proper treatment of the field products in $\mathcal{L}_{I}$ removes the ultraviolet divergences at one-loop order, in an analogous manner to the vacuum subtraction used in the standard renormalisation approach~\cite{Kapusta:2006pm,Das:1997gg,Bellac:2011kqa}. \\

\noindent
Since the relation between real and imaginary-time propagators in Eq.~\eqref{G_rel} is a general result which holds independently of the specific form of the propagators, one can repeat the analysis of Sec.~\ref{diff:ReIm} in order to express $\widetilde{\tau}_{(1), \text{TP}}^{++}(p)$ in terms of the basic thermoparticle propagators, which gives 
\begin{align}
\widetilde{\tau}_{(1), \text{TP}}^{++}(p) &= -i\left(1+\frac{1}{e^{\beta |p_{0}|}-1} \right)\widetilde{G}_{\text{TP}}(|p_{0}|+i0^{+},\vec{p})  + \frac{i}{e^{\beta |p_{0}|}-1}\widetilde{G}_{\text{TP}}(|p_{0}|-i0^{+},\vec{p}) \nonumber \\
&  + i \Delta m_{\text{TP}}^{2} \left[ \widetilde{G}_{\text{TP}}(|p_{0}|+i0^{+},\vec{p})^{2} + \frac{1}{e^{\beta |p_{0}|}-1}\left(\widetilde{G}_{\text{TP}}(|p_{0}|+i0^{+},\vec{p})^{2}-\widetilde{G}_{\text{TP}}(|p_{0}|-i0^{+},\vec{p})^{2} \right) \right].  
\label{GML_TP_2}
\end{align}
Using Eq.~\eqref{GML_TP_2} together with the spectral representation in Eq.~\eqref{rho:prop}, the corresponding one-loop spectral function $\rho_{(1), \text{TP}}(p_{0},\vec{p})$ can then be written in the form
\begin{align}
\rho_{(1), \text{TP}}(p_{0},\vec{p}) &= -i \epsilon(p_{0}) \left(\widetilde{G}_{\text{TP}}(|p_{0}|+i0^{+},\vec{p})-\widetilde{G}_{\text{TP}}(|p_{0}|-i0^{+},\vec{p}) \right)  \nonumber \\
& \quad\quad\quad\quad + i \epsilon(p_{0})\Delta m_{\text{TP}}^{2} \left(\widetilde{G}_{\text{TP}}(|p_{0}|+i0^{+},\vec{p})^{2}-\widetilde{G}_{\text{TP}}(|p_{0}|-i0^{+},\vec{p})^{2}\right)  \nonumber \\
&= \rho_{\text{TP}}(p_{0},\vec{p})\left[1 - \Delta m_{\text{TP}}^{2}\left(\widetilde{G}_{\text{TP}}(p_{0}+i0^{+},\vec{p})+\widetilde{G}_{\text{TP}}(p_{0}-i0^{+},\vec{p})\right)   \right],  
\label{rho:1-loop:Im:TP}
\end{align}
where $\rho_{\text{TP}}(p_{0},\vec{p})$ is the spectral function of the thermoparticle state. The first equality uses the property that the sum and differences of the propagators $\widetilde{G}_{\text{TP}}(|p_{0}|\pm i0^{+},\vec{p})$ as well as their squares are either purely real or imaginary, and the final equality follows from the fact that the thermoparticle propagator products are well defined irrespective of the $k_{0}\rightarrow p_{0} \pm i0^{+}$ limit, in contrast to the free-field case. Since by definition $\rho_{\text{TP}}(p_{0},\vec{p})$ has a broadened peak-like structure, Eq.~\eqref{rho:1-loop:Im:TP} implies that the effects of the thermal medium already enter at leading order in the perturbative expansion. This runs in stark contrast to the standard perturbative expansion in $\phi^{4}$ theory, where non-trivial broadening effects only start to enter at two-loop order. \\

\noindent
Another way to see the broadening dependence is on the level of the perturbative propagator itself. In the imaginary-time formalism the corresponding one-loop thermoparticle propagator is given by 
\begin{align}
\widetilde{G}_{(1), \text{TP}}(i\omega_{N},\vec{p}) = \widetilde{G}_{\text{TP}}(i\omega_{N},\vec{p}) - \Delta m_{\text{TP}}^{2} \widetilde{G}_{\text{TP}}(i\omega_{N},\vec{p})^{2},
\label{1-loop:imag:TP}
\end{align}
where now $\Delta m_{\text{TP}}^{2}$ is computed using a Matsubara sum in an analogous manner to Eq.~\eqref{deltaM_imag}, and coincides with the real-time result in Eq.~\eqref{delta_m_TP_2}. Since the analytic propagator can always be written in the general form
\begin{align}
\widetilde{G}(k_{0},\vec{p}) = -\frac{1}{k_{0}^{2}-|\vec{p}|^{2}-m^{2}-\Pi(k_{0},\vec{p})},
\end{align} 
where $\Pi(k_{0},\vec{p})$ is the corresponding self-energy, one can interpret Eq.~\eqref{1-loop:imag:TP} as the first terms in a geometric series expansion in which the one-loop propagator has the self-energy
\begin{align}
\Pi_{(1),\text{TP}}(k_{0},\vec{p})= k_{0}^{2}-|\vec{p}|^{2}-m^{2} + \widetilde{G}_{\text{TP}}(k_{0},\vec{p})^{-1} + \Delta m_{\text{TP}}^{2}.
\label{bare_SE}
\end{align} 
In the standard approach where $\widetilde{G}_{\text{TP}} \rightarrow \widetilde{G}_{(0)}$ one recovers the well-known result: $\Pi_{(1)}(k_{0},\vec{p})=\Delta m^{2}$, and the independence of $\Pi_{(1)}(k_{0},\vec{p})$ on both $k_{0}$ and $\vec{p}$ leads to a purely real shift in the propagator mass $m \rightarrow m +  \Delta m^{2}$. However, in the thermoparticle case the self-energy must necessarily have a non-trivial analytic structure due to the dependence on $\widetilde{G}_{\text{TP}}(k_{0},\vec{p})$, which is induced by the broadening effects from the damping factor in Eq.~\eqref{Matsubara:TP}. In particular, $\Pi_{(1),\text{TP}}(k_{0},\vec{p})$ possesses both real \textit{and} imaginary components, which is a realisation of the fact that the spectral function in Eq.~\eqref{rho:1-loop:Im:TP} contains a broadened component already at leading order due to the dynamical dependence of the thermoparticle scattering states.

\subsection{Perturbative spectral functions} 
\label{PT_TP_rho}

Studies of lattice $\phi^{4}$ theory~\cite{Lowdon:2024atn} are consistent at a high precision with the existence of thermoparticle states with damping factors that have a purely exponential form
\begin{align}
D_{m,\beta}(\vec{x}) = e^{-\gamma |\vec{x}|}. 
\label{exp_damping}
\end{align} 
Using the spectral representations in Eqs.~\eqref{rho_rep} and~\eqref{Matsubara:TP} it then follows that the thermoparticle spectral function and propagator are given by
\begin{align}
\rho_{\text{TP}}(p_{0},\vec{p}) &= \epsilon(p_{0})  \theta(p_{0}^{2}-m^{2}) \,  \frac{4 \gamma  \sqrt{p_{0}^{2}-m^{2}}}{(|\vec{p}|^{2}+m^{2}-p_{0}^{2})^{2} + 2(|\vec{p}|^{2}-m^{2}+p_{0}^{2})\gamma^{2}+\gamma^{4} }, \label{rho_TP_0}  \\
\widetilde{G}_{\text{TP}}(k_{0},\vec{p}) &= -\frac{1}{k_{0}^{2}-|\vec{p}|^{2}-m^{2}-2\gamma\sqrt{m^{2}-k_{0}^{2}} -\gamma^{2}}. \label{G_TP_0}
\end{align}

\begin{figure}[t!]
\centering
\includegraphics[width=0.47\textwidth]{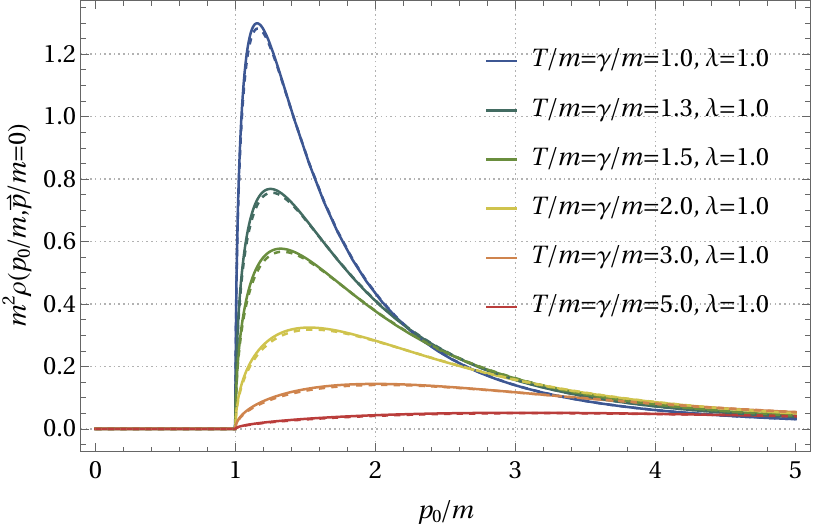}
\includegraphics[width=0.47\textwidth]{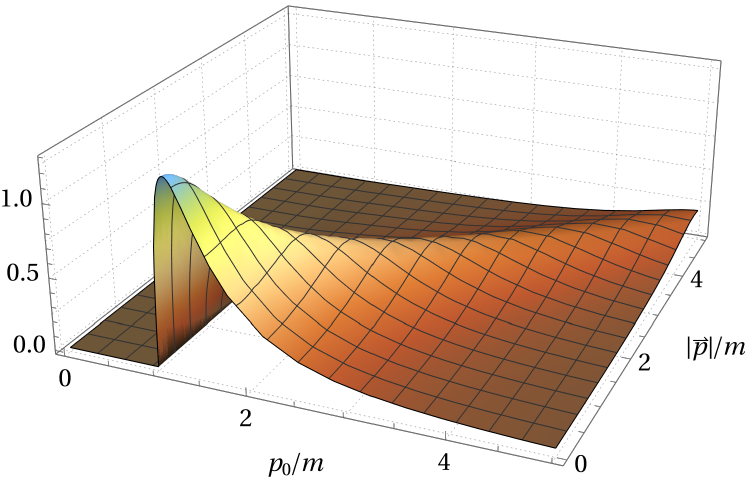}    
\caption{In the left plot is the leading order (solid lines) and one-loop (dashed lines) spectral functions at $\vec{p}=0$ for a range of $T/m$ and $\gamma/m$ values, and in the right plot is the one-loop spectral function $\rho_{(1), \text{TP}}(p_{0},\vec{p})$ at $T=\gamma=m$ as a function of both energy and momentum.}
\label{rho_plots}
\end{figure}  

\noindent
With Eqs.~\eqref{rho_TP_0} and~\eqref{G_TP_0} one can now use the one-loop perturbative result in Eq.~\eqref{rho:1-loop:Im:TP} to compute the explicit form of $\rho_{(1), \text{TP}}(p_{0},\vec{p})$ for different parameter values. The left plot of Fig.~\ref{rho_plots} shows the leading order (solid lines) and one-loop order (dashed lines) spectral functions at $\vec{p}=0$ for a range of $T/m$ and $\gamma/m$ values, and the right plot shows $\rho_{(1), \text{TP}}(p_{0},\vec{p})$ at $T=\gamma=m$ as a function of both energy and momentum. By virtue of the non-trivial damping factor both the leading and one-loop spectral functions have a broadened peak structure, in contrast to the standard result in Eq.~\eqref{rho:1-loop}, which contains only on-shell components. As discussed previously, this is a reflection of the fact that the thermoparticle scattering states are not free, and therefore the interaction effects with the medium must manifest themselves at all loop orders. The spectral function peaks increasingly broaden as a function of temperature due to the stronger interactions with the medium, and the thermoparticle threshold at $p_{0}=m$ also persists at one-loop order. At higher orders in the expansion this threshold will be lost, just as it is in the standard approach~\cite{Wang:1995qf}, since the appearance of additional thermal scattering processes as intermediate states allows for the possibility of arbitrarily small energy exchanges with the medium, so-called Landau damping.

\subsection{The thermoparticle perturbation theory framework} 
\label{framework}
  
In Sec.~\ref{PT_TP_rho} we made a specific choice for the damping factor $D_{m,\beta}(\vec{x})$ in order illustrate the main differences between thermoparticle perturbation theory and the standard approach. However, as outlined in Sec.~\ref{TP}, the damping factor should actually be determined by the dynamics of the theory in question. In particular, in Ref.~\cite{Bros:2001zs} the authors proposed a consistency condition for the asymptotic thermoparticle fields $\phi_{\text{TP}}(x)$, and in the case of $\phi^{4}$ theory demonstrated that this constraint uniquely fixes the form of $D_{m,\beta}(\vec{x})$. Together with the modified GML relation Eq.~\eqref{GML_TP}, this condition sets a procedure by which one can in principle perform thermoparticle perturbation theory calculations for any finite-temperature QFT:

\begin{enumerate}[label=\bf{Step\,\arabic{enumi}.}, wide=0pt, leftmargin=*]

\item Apply the asymptotic field condition to compute the damping factor $D_{m,\beta}(\vec{x})$ of the corresponding thermoparticle fields $\phi_{\text{TP}}(x)$. 

\item Use $D_{m,\beta}(\vec{x})$ to derive the thermoparticle propagator, and perform an expansion of the GML relation in Eq.~\eqref{GML_TP} to compute the perturbative prediction up to some fixed loop order.
 
\end{enumerate}

It should be noted that the first step constitutes a non-perturbative problem, since the damping factor necessarily involves a non-vanishing coupling. This is a fundamental difference to the standard perturbative approach, and arises due to the impossibility of realising free scattering states in a thermal medium. \\

\noindent
As a starting point we have focussed in this work on QFTs containing a single species of scalar field $\phi(x)$, but the thermoparticle framework can certainly be generalised to QFTs with fields of higher spin, and with a larger field content, which would be important for understanding more physically sophisticated theories such as QED or QCD. We leave such extensions, as well as a systematic investigation of the higher-order renormalisation procedure, to future work. In the remainder of this work we will demonstrate the practical feasibility of thermoparticle perturbation theory.

\section{Lattice test of thermoparticle perturbation theory}
\label{lattice_test}

\subsection{Lattice formulation}

In Ref.~\cite{Lowdon:2024atn} perturbative predictions of correlation functions in $\phi^{4}$ theory using the standard finite-temperature approach were compared with lattice simulation data, and it was demonstrated that these predictions increasingly break down for higher temperatures\footnote{See Ref.~\cite{Romatschke:2026tam} for another recent study on the structural properties of $\phi^{4}$ theory, both in vacuum and at finite temperature.}. In this section we test the lattice-adapted thermoparticle perturbation theory framework against the standard approach and simulation data using the same setup as in Ref.~\cite{Lowdon:2024atn}. The lattice action of the theory is defined by
\begin{align}
S = a^{4}\sum_{x \in \Lambda_{a}} \left[ \frac{1}{2}\sum_{\mu} \Delta_{\mu}^{f}\phi(x)\Delta_{\mu}^{f}\phi(x)+ \frac{m_{0}^{2}}{2}\phi(x)^{2} +\frac{g_{0}}{4!}\phi(x)^{4} \right], 
\label{eq:latact1}
\end{align}
where $\Delta_{\mu}^{f}$ is the lattice forward derivative: $\Delta_{\mu}^{f}\phi(x)= \left[\phi(x+a\hat{n}_{\mu})-\phi(x)\right]/a$. The field $\phi(x)$ is taken to be periodic in each direction, with the temporal periodicity implying that the system has a corresponding temperature $T=1/(aN_{\tau})$. The zero-temperature QFT at fixed lattice spacing and spatial volume $V=(aN_{s})^{3}$ is then recovered in the limit $N_{\tau}\rightarrow\infty$. In this study we focus on the symmetric phase with positive bare masses $m_{0}$. By keeping the lattice spacing $a>0$ small but fixed, this avoids the potential triviality of the theory in the continuum ($a\rightarrow 0$) limit, and ensures that the theory behaves like an interacting effective theory with negligible discretisation effects~\cite{Montvay:1987us,Montvay:1994cy}. \\

\noindent
The observable of interest in this study is the Euclidean two-point correlation function $\langle \phi(\tau,\vec{x})\phi(0)\rangle$. In particular, we focus on the spatial and temporal correlators, which on the lattice are defined as the following discrete sums over the orthogonal spacetime directions\footnote{For ease of notation we have suppressed the explicit functional dependence on the spacing $a>0$ in the argument of these lattice correlators, as well as those discussed in the remainder of this section.} 
\begin{align}
C(z;N_{s},N_{\tau}) &=  a^{3}\sum_{\tau,x,y}\langle \phi(\tau,\vec{x})\phi(0)\rangle, \label{latt_Cz} \\
C(\tau;N_{s},N_{\tau}) &=  a^{3}\sum_{x,y,z}\langle \phi(\tau,\vec{x})\phi(0)\rangle. \label{latt_Ct}
\end{align}
The spatial and temporal correlators encode complementary information about the spectral properties of the thermal medium, and both quantities can be straight-forwardly computed in perturbation theory as well as by numerical simulations. \\

\noindent
Following Ref.~\cite{Lowdon:2024atn}, we use lattice perturbation theory since its predictions can be directly compared with simulated data without needing to take into account finite-volume or discretisation effects, which are the same in both cases. Choosing identical bare parameter sets for the perturbative and numerical predictions thus ensures that any statistically significant differences between calculations are entirely due to the perturbative approximation itself. The free-field lattice propagator is
\begin{align}
\widetilde{G}_{0}(p;N_{s},N_{\tau}) = \frac{1}{\sum_{\mu}\tfrac{4}{a^{2}}\sin^{2}\left(\tfrac{a p_{\mu}}{2}\right)+m_{0}^{2}},
\label{prop_free}
\end{align}
where the momenta $p_{\mu}$ are restricted to the Brillouin zone $\mathcal{B}_{a}= \{p_{\mu}= \tfrac{2\pi }{aN_{\mu}}k_{\mu}, \, k_{\mu}= 0, \dots, N_{\mu}-1 \}$, and $p_{\tau}$ is the energy conjugate to Euclidean time $\tau$. For an infinite lattice $N_{s}=N_{\tau}\rightarrow \infty$, the standard continuum Euclidean propagator $(p_{\tau}^{2}+ |\vec{p}|^{2}+m_0^{2})^{-1}$ is recovered in the $a\rightarrow 0$ limit. By defining the self-energy $\Pi(p;N_{s},N_{\tau})$ as the sum of all one-particle irreducible diagrams, as in the continuum, the full lattice-discretised propagator of the interacting theory $\widetilde{G}(p;N_{s},N_{\tau})$ corresponds to a geometrical series in $\Pi$, which can be summed to
\begin{align}
\widetilde{G}(p;N_{s},N_{\tau}) =  \frac{1}{\sum_{\mu}\tfrac{4}{a^{2}}\sin^{2}\left(\tfrac{a p_{\mu}}{2}\right)+m_{0}^{2} + \Pi(p;N_{s},N_{\tau})}.
\label{prop_full} 
\end{align}
Given the self-energy $\Pi$ to some order, Eq.~\eqref{prop_full} together with the definitions in Eqs.~\eqref{latt_Cz} and~\eqref{latt_Ct} provides the lattice perturbation theory predictions for the spatial and temporal correlators
\begin{align}
C(z;N_{s},N_{\tau}) &=  \frac{1}{N_{s}} \! \sum_{k_{z}=0}^{N_{s}-1} \! e^{\frac{2\pi i k_{z}}{aN_{s}}z}  \frac{a}{4\sin^{2}\!\left(\tfrac{\pi k_{z} }{N_{s}}\right)+(am_{0})^{2}+a^{2}\Pi(\omega_{E}=p_{x}=p_{y}=0,p_{z}=\frac{2\pi k_{z}}{aN_{s}};N_{s},N_{\tau})}, \label{Cz} \\  
C(\tau;N_{s},N_{\tau}) &= \frac{1}{N_{\tau}} \! \sum_{k_{\tau}=0}^{N_{\tau}-1} \! e^{\frac{2\pi i k_{\tau}}{aN_{\tau}}\tau} \frac{a}{4\sin^{2}\!\left(\tfrac{\pi k_{\tau} }{N_{\tau}}\right)+(am_{0})^{2}+a^{2}\Pi(k_{0}=\frac{2\pi i k_{\tau}}{aN_{\tau}},p_{x}=p_{y}=p_{z}=0;N_{s},N_{\tau})}.  
\label{Ct}
\end{align}

\begin{figure}[t!] 
\centering
\includegraphics[width=0.65\textwidth]{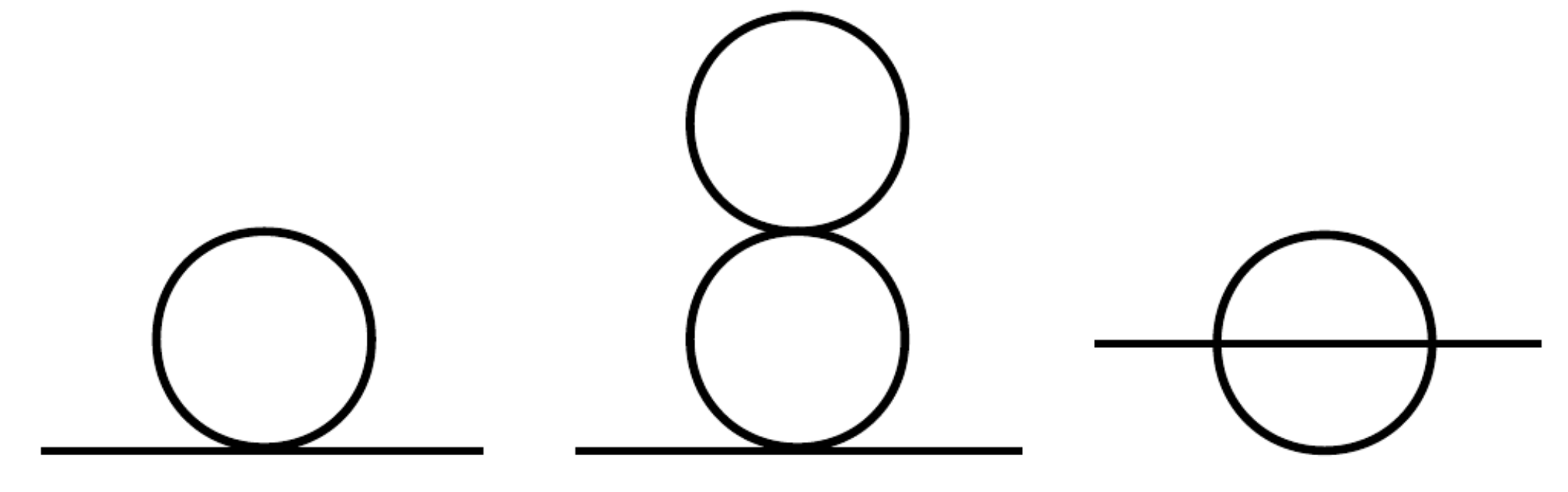}
\caption{The contributing diagrams to the self-energy up to $\mathcal{O}(g_{0}^{2})$. From left to right these are the bubble, cactus, and sunset diagrams, respectively.}
\label{self_energy_diag}
\end{figure}

In the following we compute the self energy up to $\mathcal{O}(g_{0}^{2})$ (two-loop order), using the Feynman diagrams displayed in Fig.~\ref{self_energy_diag}, which leads to the explicit form~\cite{Montvay:1994cy} 
\begin{align}
a^{2}\Pi_{(2)}(p;N_{s},N_{\tau}) &= \frac{g_{0}}{2}J_{1}(am_{0};N_{s},N_{\tau})  -\frac{g_{0}^{2}}{6}I_{3}(p,am_{0}; N_{s},N_{\tau}) \nonumber \\
&\quad\quad\quad\quad\quad\quad -\frac{g_{0}^{2}}{4}J_{1}(am_{0}; N_{s},N_{\tau}) \, J_{2}(am_{0}; N_{s},N_{\tau}),  \label{2loop_SE} \\
J_{n}(am_{0};N_{s},N_{\tau}) &= \frac{1}{N_{s}^{3}N_{\tau}}\sum_{p \in \mathcal{B}_{a}} \widetilde{G}_{0}(p;N_{s},N_{\tau})^{n}, \quad\quad n=1,2 \label{Jn} \\
I_{3}(p,am_{0};N_{s},N_{\tau}) &= \frac{1}{(N_{s}^{3}N_{\tau})^{2}} \sum_{q\in \mathcal{B}_{a}}  \sum_{r\in \mathcal{B}_{a}} \widetilde{G}_{0}(p-q-r;N_{s},N_{\tau}) \, \widetilde{G}_{0}(q;N_{s},N_{\tau}) \, \widetilde{G}_{0}(r;N_{s},N_{\tau}). \label{I3}
\end{align}
In order to maximise sensitivity to finite-temperature effects we choose the bare parameters $(am_{0},g_{0})$ such that the two-loop perturbative predictions are sufficient to describe the simulation data on the vacuum-like large-$N_{\tau}$ lattices, and with smaller values of $N_{\tau}$ showing significant changes. We consider only moderate lattice sizes of $16^{3}\times N_{\tau}$ with $2 \leq N_{\tau} \leq 16$ to keep the sums in Eq.~\eqref{I3} manageable, and find that in this case the choice $(am_{0}=0.2,g_{0}=1.298)$ meets our criteria. \\

\begin{figure}[t!]
\centering
\includegraphics[width=0.43\textwidth]{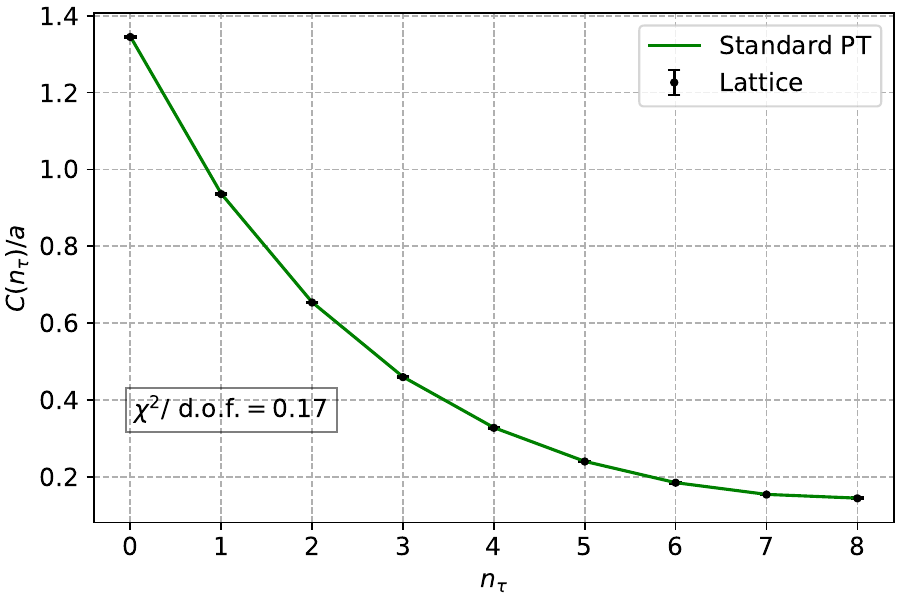}
\includegraphics[width=0.43\textwidth]{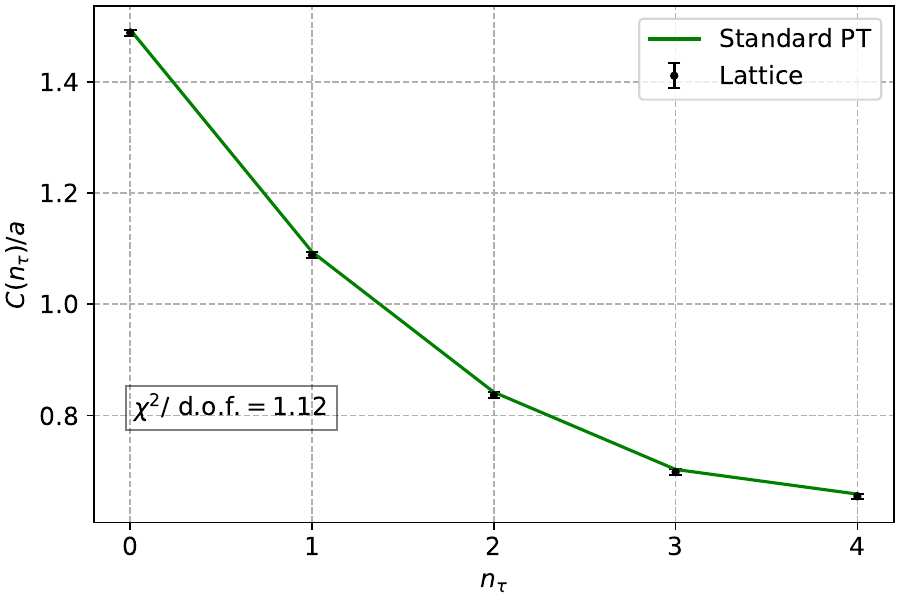}
\includegraphics[width=0.43\textwidth]{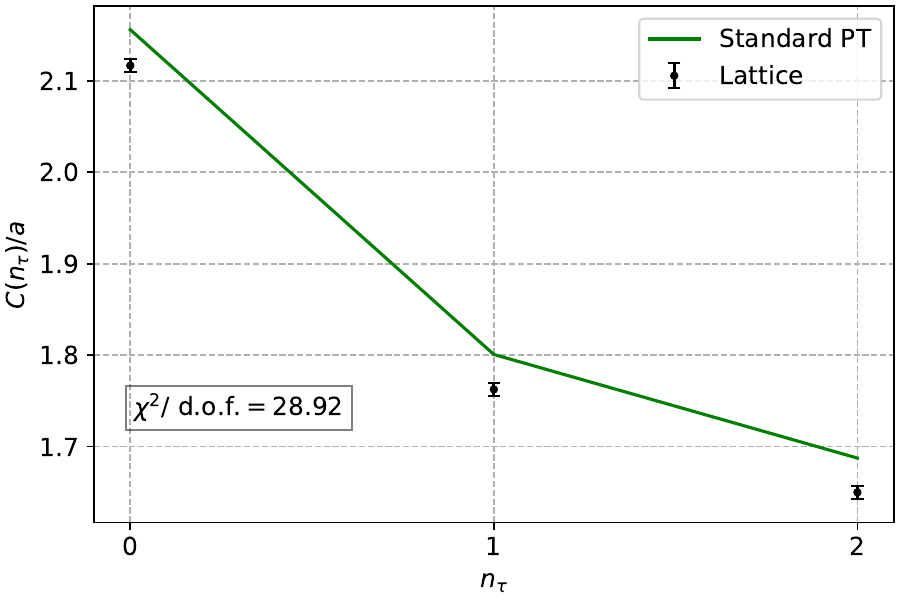}
\includegraphics[width=0.43\textwidth]{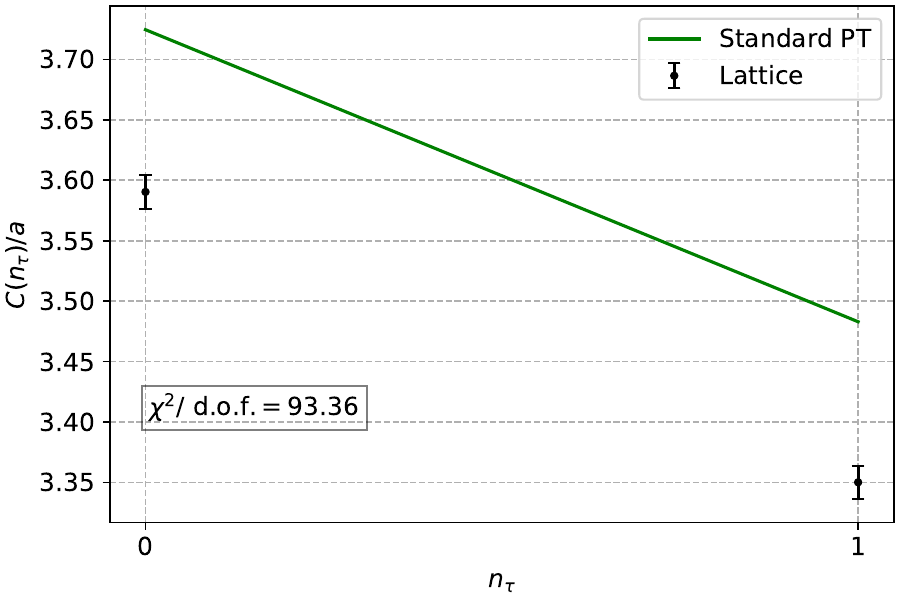}
\caption{Comparison of the two-loop perturbative predictions for the temporal correlator in standard perturbation theory (PT) with the corresponding $16^3\times N_{\tau}$ lattice data (black points) at $N_{\tau}=16,8,4,2$ (top left, top right, bottom left, bottom right) for $(am_{0}=0.2,g_{0}=1.298)$. The predictions at different temporal points are joined together for ease of comparison.}
\label{vac_temporal_am02_g1298}
\end{figure}

\noindent
Fig.~\ref{vac_temporal_am02_g1298} shows the corresponding two-loop temporal correlator predictions. They are quantitatively consistent with the simulation data of the coldest system with $N_{\tau}=16$. However, for smaller temporal lattice sizes ($N_{\tau}=8,4,2$) the predictions start to increasingly deteriorate\footnote{As in Ref.~\cite{Lowdon:2024atn}, the reduced chi-squared $\chi^{2}/\text{d.o.f.}$ values of the perturbative predictions are computed in order to provide a quantitative measure of how well the predictions describe the data over the full lattice range. These values are displayed in Fig.~\ref{vac_temporal_am02_g1298}, as well as in subsequent temporal correlator prediction plots in Secs.~\ref{latt_TP_PT} and~\ref{prop_sens}.}. These deviations at smaller values of $N_{\tau}$ must therefore be due to temperature effects alone, and are not simply a result of missing higher-order corrections. This demonstrates that even for massive theories, in which there are no explicit infrared divergences, the standard finite-temperature perturbative procedure breaks down, as was shown previously in Ref.~\cite{Lowdon:2024atn}. The origin of these deviations is the infrared behaviour of the free-field propagator in Eq.~\eqref{prop_free}, namely that the $p_{\mu}=0$ contribution is highly sensitive to the value of $am_{0}$. As $am_{0}$ becomes smaller for fixed $g_{0}$, these zero-mode contributions grow unbounded, causing the predictions to deviate significantly from the lattice data, which is simply a realisation of the infrared problem detailed in Sec.~\ref{probs}. 

\subsection{Thermoparticle perturbation theory}
\label{latt_TP_PT}

In the remainder of this section we explore the practical application of our proposed new framework. As outlined in Sec.~\ref{framework}, the main difficulty in thermoparticle perturbation theory is to determine the specific form of the damping factor $D_{m,\beta}(\vec{x})$ of the thermoparticle states. In Ref.~\cite{Bros:2001zs} the authors explicitly derived $D_{m,\beta}(\vec{x})$ in $\phi^{4}$ theory for both positive and negative coupling using the consistency condition in Eq.~\eqref{asymptot_cond}. However, in the positive coupling case, which is the focus of our study, the authors had to renormalise the fields in order to ensure the existence of a meaningful solution for $D_{m,\beta}(\vec{x})$, which is related to the fact that the theory is most-likely trivial in the continuum limit. Since Eq.~\eqref{asymptot_cond} is defined for continuum theories, it is not immediately clear how one should compute $D_{m,\beta}(\vec{x})$ in lattice-regularised $\phi^{4}$ theory. An alternative strategy was pointed out in Ref.~\cite{Lowdon:2022xcl}, where the functional form of $D_{m,\beta}(\vec{x})$ was determined from spatial correlator data itself. In particular, real $\phi^{4}$ theory was shown at high precision to be consistent with having a purely exponential damping factor. Given that $D_{m,\beta}(\vec{x}) = e^{-\gamma_{0} |\vec{x}|}$, it immediately follows from Eq.~\eqref{G_TP_0} that the corresponding lattice-regularised thermoparticle propagator has the form 
\begin{align}
\widetilde{G}_{\text{TP}}(p;N_{s},N_{\tau}) &= \frac{1}{\sum_{\mu}\tfrac{4}{a^{2}}\sin^{2}\left(\tfrac{a p_{\mu}}{2}\right)+m_{0}^{2}+2\gamma_{0}\sqrt{m_{0}^{2}+\tfrac{4}{a^{2}}\sin^{2}\left(\tfrac{a p_{\tau}}{2}\right)} +\gamma_{0}^{2}}, \label{G_TP_phi4}
\end{align}
where now $\gamma_{0}=\gamma_{0}(m_{0},g_{0},N_{s},N_{\tau})$ is a function of all the lattice parameters. In Sec.~\ref{GML_rel_TP} it was shown that the topology of the diagrams in thermoparticle perturbation theory is identical to the standard approach. The corresponding two-loop predictions of the spatial and temporal correlators can thus be obtained using Eqs.~\eqref{2loop_SE}-\eqref{I3} with the replacement: $\widetilde{G}_{0} \rightarrow \widetilde{G}_{\text{TP}}$. \\

\noindent
The only remaining issue is to determine the value of the dimensionless damping parameter $a\gamma_{0}$ for a given choice of lattice parameters, and make predictions based on it. We do so by performing the following steps:

\begin{enumerate}[label=\bf{\arabic{enumi}.}]

\item Compute the two-loop perturbative prediction for the spatial correlator $C_{\text{TP}}(z,a\gamma_{0})$ using a sequence of different values for $a\gamma_{0}$.

\item Identify the values of $a\gamma_{0}$ for which the perturbative predictions are consistent with the simulated spatial correlator data. 

\item Use propagators with these $a\gamma_{0}$ values to predict other observables.
    
\end{enumerate} 
 
\begin{table}[t!]  
\center
\small
\renewcommand{\arraystretch}{1.35}
\begin{tabular}{|c|c|c|} 
\hline
\rule{0pt}{3ex}
$N_{s}^{3} \times N_{\tau}$ & $a\gamma_{0}^{*}{}^{+\sigma_{\text{max}}}_{-\sigma_{\text{min}}}$   & $\chi^{2}/\text{d.o.f.}$   \\[0.5ex]
\hhline{|=|=|=|}
$16^{3} \times 8$    & $0.0020{}^{+0.0015}_{-0.0020}$  & 0.03  \\
\hline
$16^{3} \times 4$    & $0.0070{}^{+0.0010}_{-0.0016}$  & 0.08  \\
\hline
$16^{3} \times 2$    & $0.0110{}^{+0.0013}_{-0.0011}$  & 0.07   \\ 
\hline
\end{tabular}
\caption{Tuned $a\gamma_{0}$ values obtained by matching the two-loop thermoparticle spatial correlator prediction $C_{\text{TP}}(z, a\gamma_{0},N_{\tau})$ to the corresponding lattice data. $a\gamma_{0}^{*}$ corresponds to the $a\gamma_{0}$ value found with the lowest reduced $\chi^{2}$ prediction, and $\sigma_{\text{max/min}}$ are the upper/lower ranges of these predictions beyond which $\chi^{2}/\text{d.o.f.} \gtrsim 1$.}
\label{tab:gamma0_Nt}
\end{table} 

\noindent
The tuning procedure in step 2 constitutes a significant initial test of the thermoparticle approach since the spatial correlator predictions are fixed by a single parameter $a\gamma_{0}$. We found that for each value of $N_{\tau}$ this parameter could be tuned to a value $a\gamma_{0}^{*}$ such that the $C_{\text{TP}}(z, a\gamma_{0},N_{\tau})$ predictions gave a consistent description of the simulated data over the entire spatial correlation range $0 \leq z \leq (aN_{s})/2$. We also varied $a\gamma_{0}$ above and below $a\gamma_{0}^{*}$ until the predictions started to deteriorate with $\chi^{2}/\text{d.o.f.} \gtrsim 1$, resulting in an effective consistency band $[a\gamma_{0}^{*}-\sigma_{\min},a\gamma_{0}^{*}+\sigma_{\max}]$. These values together with the reduced $\chi^{2}$ of the spatial correlator predictions are displayed in Table~\ref{tab:gamma0_Nt} for $N_{\tau}=8,4,2$.  \\

\begin{figure}[t!] 
\centering 
\includegraphics[width=0.43\textwidth]{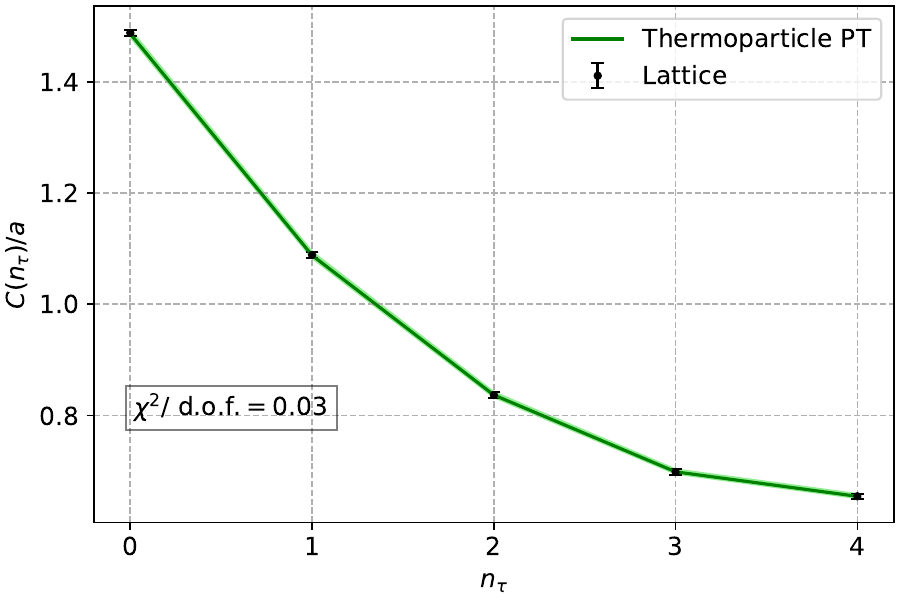}
\includegraphics[width=0.43\textwidth]{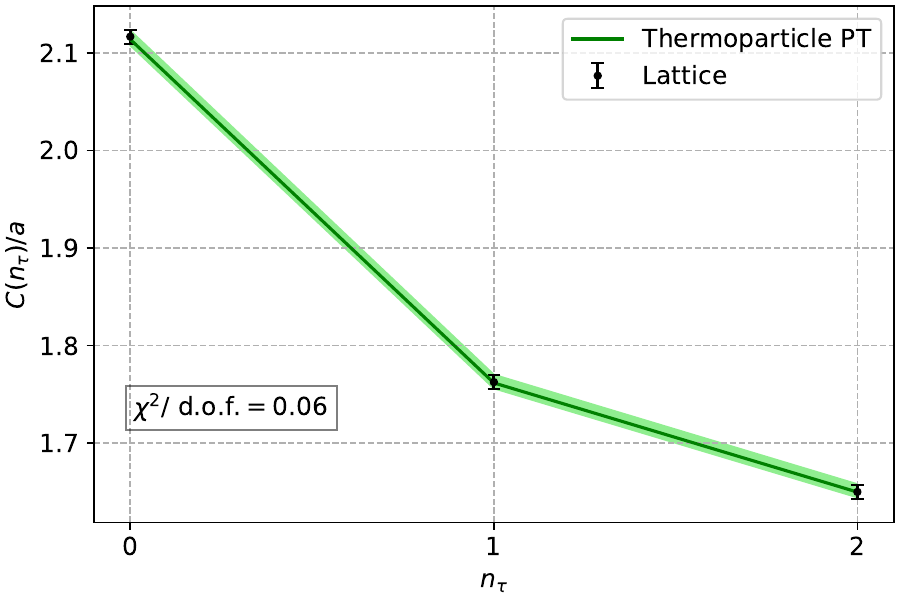}
\includegraphics[width=0.43\textwidth]{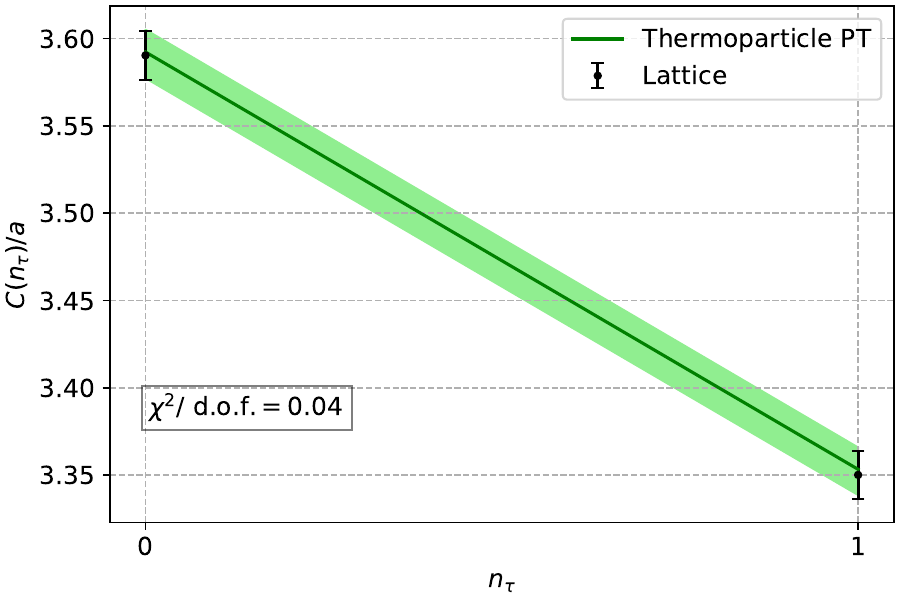}
\caption{Comparison of the two-loop thermoparticle perturbation theory (PT) predictions for the temporal correlator with the corresponding $16^3\times N_{\tau}$ lattice data (black points) at $N_{\tau}=8,4,2$ (top left, top right, bottom) for $(am_{0}=0.2,g_{0}=1.298)$. The predictions at different temporal points are joined together for ease of comparison, and the bands correspond to the effective uncertainties on these predictions.}
\label{TP_temporal_am02_g1298}
\end{figure}

\begin{figure}[t!]
\centering 
\includegraphics[width=0.43\textwidth]{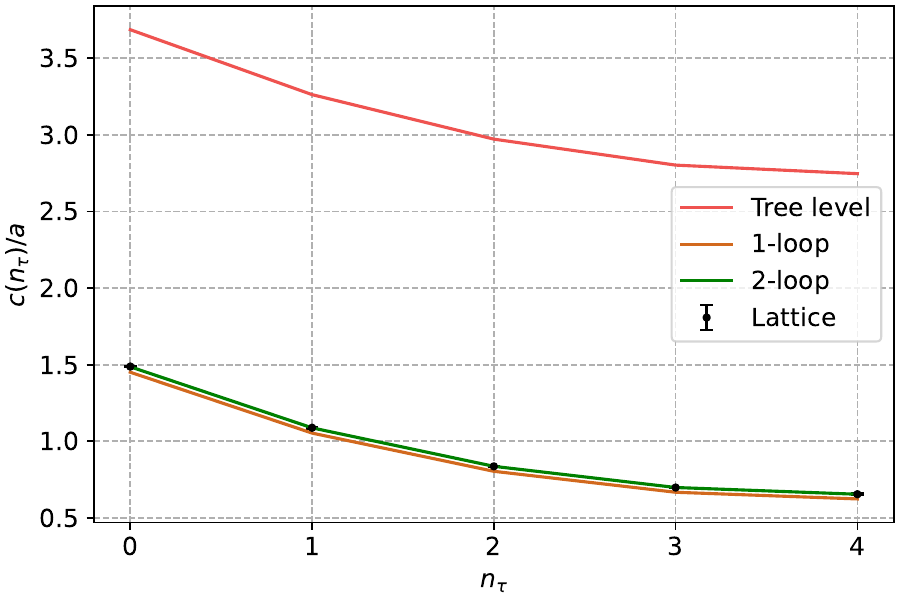}
\includegraphics[width=0.43\textwidth]{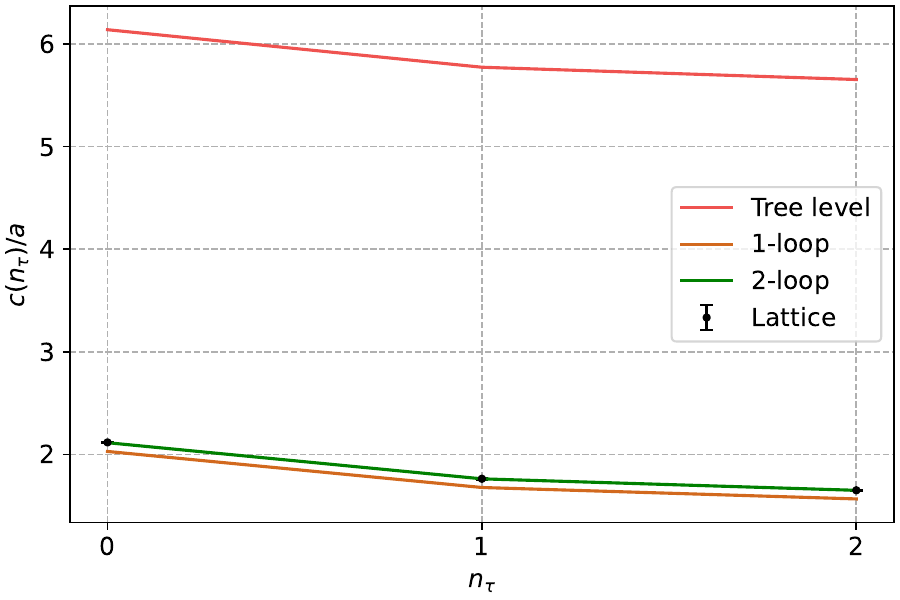}
\includegraphics[width=0.43\textwidth]{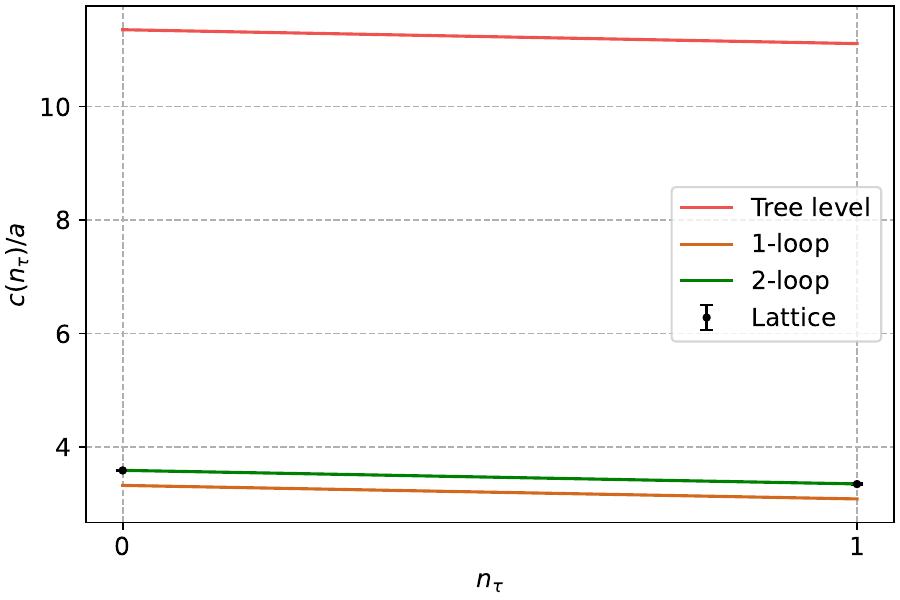} 
\caption{Comparison of the tree-level, one-loop, and two-loop thermoparticle perturbation theory predictions of the temporal correlator with the corresponding $16^3\times N_{\tau}$ lattice data (black points) at $N_{\tau}=8,4,2$ (top left, top right, bottom) for $(am_{0}=0.2,g_{0}=1.298)$. The predictions at different temporal points are joined together for ease of comparison.}
\label{TP_asym_conv} 
\end{figure}

\noindent
In step 3 we focus on the predictions of the two-loop temporal correlator $C_{\text{TP}}(\tau, a\gamma_{0},N_{\tau})$, which we compare with the corresponding lattice data. Since any consistent perturbative description must be able to simultaneously describe both the spatial \textit{and} temporal correlation functions, this prediction provides a highly non-trivial test of the validity of thermoparticle perturbation theory. The results are displayed in Fig.~\ref{TP_temporal_am02_g1298}, together with the corresponding lattice data for $N_{\tau}=8,4,2$. For $N_{\tau}=16$ the data is consistent with $a\gamma_{0}=0$, which coincides with the standard approach displayed in Fig.~\ref{vac_temporal_am02_g1298}. The bands in Fig.~\ref{TP_temporal_am02_g1298} correspond to the uncertainties on these predictions arising from the range of potentially consistent $a\gamma_{0}$ values within $[a\gamma_{0}^{*}-\sigma_{\min},a\gamma_{0}^{*}+\sigma_{\max}]$. In stark contrast to the standard approach in Fig.~\ref{vac_temporal_am02_g1298}, the thermoparticle perturbative predictions are consistent with the lattice data within errors at each value of $N_{\tau}$. This is quite remarkable given the very small statistical errors (less than $1\%$). In Fig.~\ref{TP_asym_conv} we also plot the tree-level, one-loop, and two-loop results for the temporal correlator predictions using the $a\gamma_{0}^{*}$ values for $N_{\tau}=8,4,2$. One can see that the thermoparticle predictions get successively closer to the data at higher orders, which indicates that the perturbative series has an asymptotic-like convergence behaviour, as in the vacuum theory. Analogous figures in Ref.~\cite{Lowdon:2022xcl} show that this is clearly not the case for the standard finite-temperature perturbative approach, which is further evidence that the incorrect choice of free-field scattering states in the standard approach is the source of its poor convergence properties.

\subsection{Sensitivity to the propagator structure}
\label{prop_sens} 

The high-precision results from the last section are good evidence that thermoparticles are the appropriate thermal scattering states. But how unique are these predictions? To answer this question we investigate a different causal parametrisation, namely the relativistic Breit-Wigner, whose spectral function is given in Eq.~\eqref{rho_BW:p}. This is of particular interest since it has been proposed as a possible parametrisation in previous attempts to generalise perturbation theory to finite temperatures, as outlined in Sec.~\eqref{origin}. Given the spectral function in Eq.~\eqref{rho_BW:p} it follows that the corresponding Euclidean lattice-regularised propagator has the form
\begin{align}
\widetilde{G}_{\text{BW}}(p;N_{s},N_{\tau}) &= \frac{1}{\sum_{\mu}\tfrac{4}{a^{2}}\sin^{2}\left(\tfrac{a p_{\mu}}{2}\right)+m_{0}^{2}+2\Gamma_{0} \left|\tfrac{2}{a}\sin\left(\tfrac{a p_{\tau}}{2}\right) \right| + \Gamma_{0}^{2}}. \label{G_BW_phi4}
\end{align}
The corresponding two-loop calculations of the spatial and temporal correlators can once again be computed using Eqs.~\eqref{2loop_SE}-\eqref{I3}, but now with the replacement $\widetilde{G}_{0}\rightarrow \widetilde{G}_{\text{BW}}$. As in the thermoparticle case, we fix the value of the parameter $a\Gamma_{0}$ by matching the two-loop spatial correlator prediction $C_{\text{BW}}(z, a\Gamma_{0},N_{\tau})$ to the simulation data. The parameters $a\Gamma_{0}^{*}$ together with their consistency bands and spatial correlator reduced $\chi^{2}$ values are displayed in Table~\ref{tab:Gamma0_Nt}. \\

\begin{table}[t!]
\center
\small
\renewcommand{\arraystretch}{1.35}
\begin{tabular}{|c|c|c|} 
\hline
\rule{0pt}{3ex}
$N_{s}^{3} \times N_{\tau}$ & $a\Gamma_{0}^{*}{}^{+\sigma_{\text{max}}}_{-\sigma_{\text{min}}}$   & $\chi^{2}/\text{d.o.f.}$   \\[0.5ex]
\hhline{|=|=|=|}
$16^{3} \times 8$    & $0.0450{}^{+0.0100}_{-0.0150}$  & 0.02  \\
\hline
$16^{3} \times 4$    & $0.0650{}^{+0.0035}_{-0.0060}$  & 0.07  \\
\hline
$16^{3} \times 2$    & $0.0750{}^{+0.0015}_{-0.0060}$  & 0.37   \\ 
\hline
\end{tabular}
\caption{Tuned $a\Gamma_{0}$ values obtained by matching the two-loop Breit-Wigner spatial correlator prediction $C_{\text{BW}}(z, a\Gamma_{0},N_{\tau})$ to the corresponding lattice data. $a\Gamma_{0}^{*}$ corresponds to the $a\Gamma_{0}$ value found with the lowest reduced $\chi^{2}$ prediction, and $\sigma_{\text{max/min}}$ are the upper/lower ranges of these predictions beyond which $\chi^{2}/\text{d.o.f.} \gtrsim 1$.}
\label{tab:Gamma0_Nt}
\end{table} 

\begin{figure}[t!]
\centering
\includegraphics[width=0.43\textwidth]{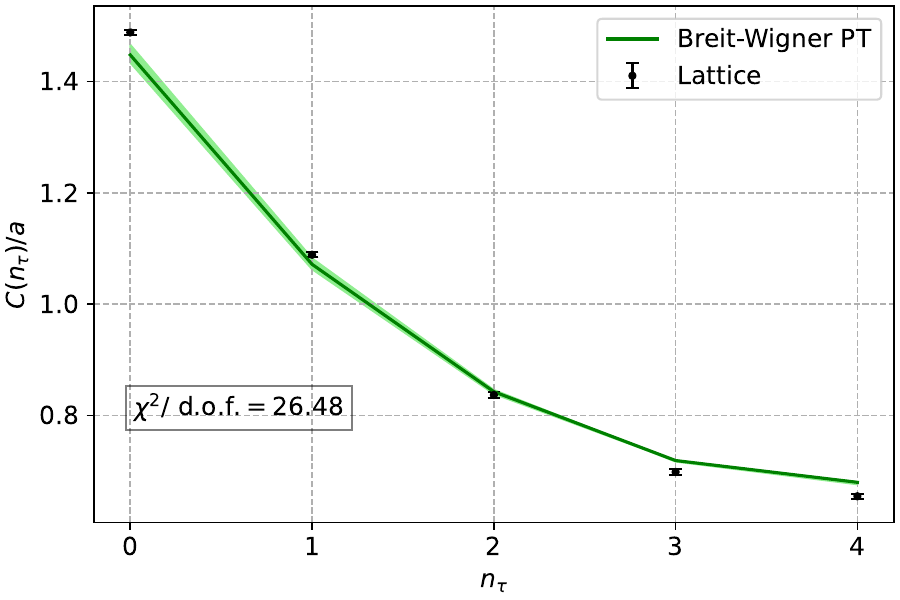}
\includegraphics[width=0.43\textwidth]{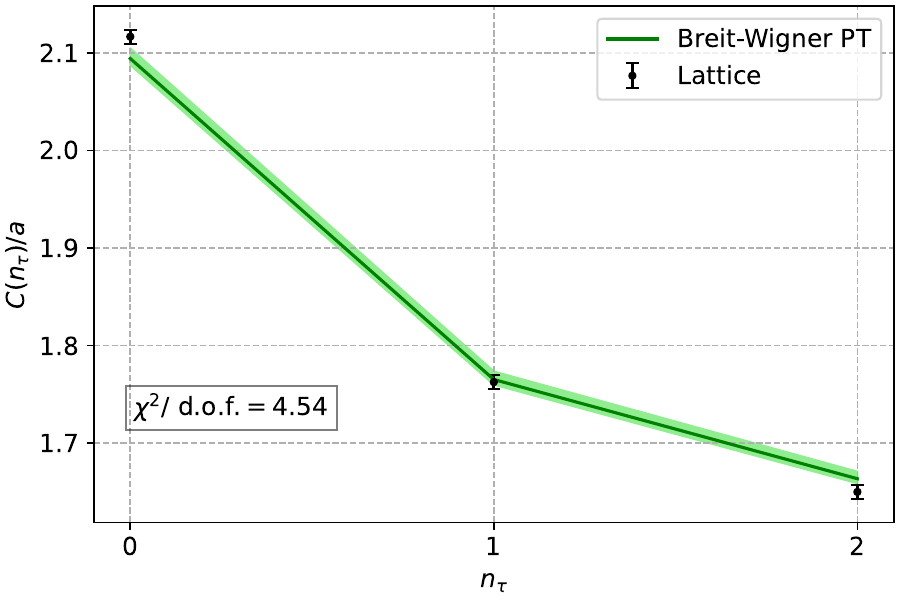} 
\includegraphics[width=0.43\textwidth]{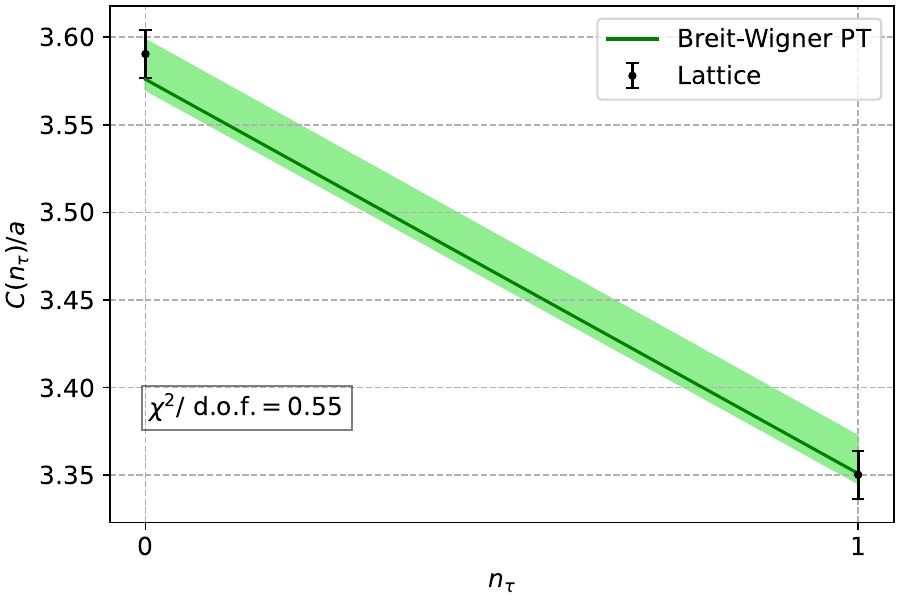}
\caption{Comparison of the two-loop Breit-Wigner perturbation theory (PT) predictions for the temporal correlator with the corresponding $16^3\times N_{\tau}$ lattice data (black points) at $N_{\tau}=8,4,2$ (top left, top right, bottom) for $(am_{0}=0.2,g_{0}=1.298)$. The predictions at different temporal points are joined together for ease of comparison, and the bands correspond to the effective uncertainties on these predictions.}
\label{BW_temporal_am02_g1298}
\end{figure}

\noindent
The resulting temporal correlator predictions are shown in Fig.~\ref{BW_temporal_am02_g1298} for $N_{\tau}=8,4,2$. For $N_{\tau}=16$ the data is consistent with $a\Gamma_{0}=0$, which coincides with the standard approach displayed in Fig.~\ref{vac_temporal_am02_g1298}. For $N_{\tau}=4$ and 8 there are significant deviations from the data, and for $N_{\tau}=2$ there is reasonable agreement, although we observe some tension which would most likely increase with smaller statistical errors on the temporal correlator data. The perturbative description using Breit-Wigner-type propagators is closer to the data for larger temperatures compared with the standard perturbative case in Fig.~\ref{vac_temporal_am02_g1298}, but deviates more significantly for lower temperatures. This is due to the fact that the Breit-Wigner propagators are no longer as sensitive to zero-mode contributions. However, for all values of $N_{\tau}$ these predictions are significantly worse than in the thermoparticle case. These results demonstrate that the perturbative predictions are highly sensitive to the functional form of the field propagators used in the expansion, and ultimately indicate that the finite-temperature scattering states are not consistent with having the Breit-Wigner form proposed in Ref.~\cite{Landsman:1988ta}.

\subsection{Parametric dependence of the damping parameter}
\label{gamma0_dep}

\begin{figure}[t!]
\centering
\includegraphics[width=0.43\textwidth]{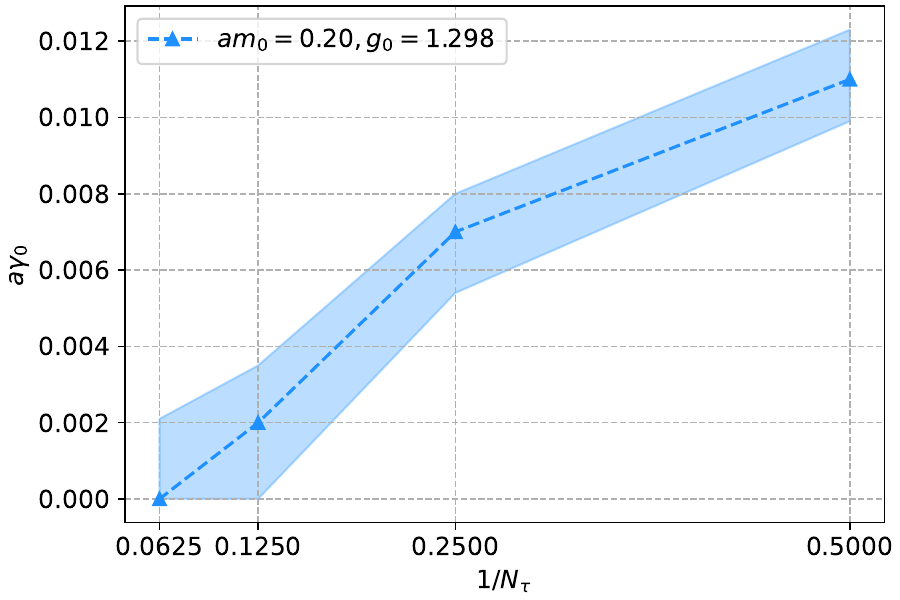}
\includegraphics[width=0.43\textwidth]{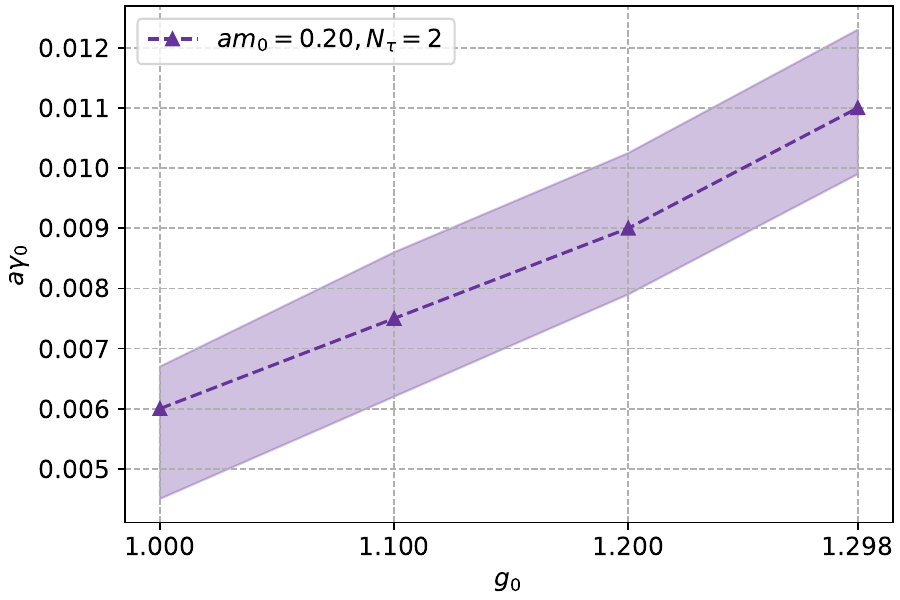}
\caption{$a\gamma_{0}$ and its uncertainty as a function of $1/N_{\tau}$ for $am_{0}=0.2$, $g_{0}=1.298$ (left), and as a function of $g_{0}$ for $am_{0}=0.2$, $N_{\tau}=2$ (right). The $a\gamma_{0}$ values for different $1/N_{\tau}$ (left) and $g_{0}$ (right) are joined together for ease of comparison.}
\label{TP_Nt_g0}
\end{figure}

\begin{figure}[t!]
\centering
\includegraphics[width=0.43\textwidth]{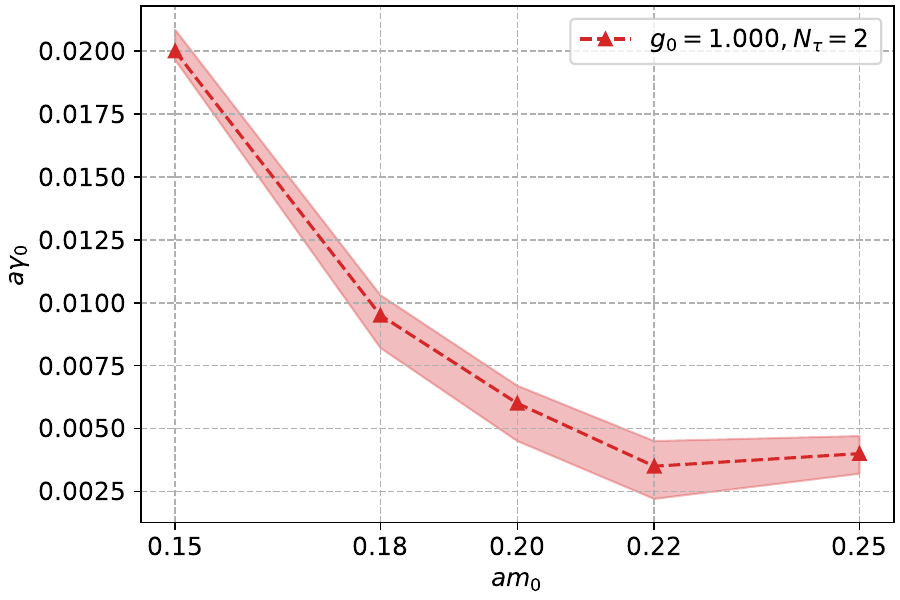}
\caption{$a\gamma_{0}$ and its uncertainty as a function of $am_{0}$ for $g_{0}=1.0$, $N_{\tau}=2$. The $a\gamma_{0}$ values at different $am_{0}$ are joined together for ease of comparison.}
\label{TP_am0}
\end{figure}

\noindent
Since the thermoparticle predictions provide a consistent description of all of the available lattice data, one can further explore the associated thermal physics by establishing how the dimensionless damping parameter $a\gamma_{0}(am_{0},g_{0},N_{s},N_{\tau})$ depends on the lattice parameters. To do so, we vary the temperature, the bare mass $am_{0}$, and the coupling strength $g_{0}$, and in each case tune the damping parameter so as to achieve consistency with the simulation results, as before. The left plot of Fig.~\ref{TP_Nt_g0} shows the corresponding $a\gamma_{0}^{*}$ value and its consistency band as a function of $aT=1/N_{\tau}$ for $am_{0}=0.2$, $g_{0}=1.298$. One can see that $a\gamma_{0}$ grows with increasing $1/N_{\tau}$, and rapidly approaches zero in the small $1/N_{\tau}$ regime. The growth of $a\gamma_{0}$ with temperature is entirely expected since $a\gamma_{0}$ captures the medium effects experienced by the asymptotic states, which are more pronounced at higher temperatures. This is also consistent with the analytic continuum investigations in Ref.~\cite{Bros:2001zs}, where the width parameter in the damping factor was found to grow linearly at large $T$ and decrease exponentially at small $T$. The right plot of Fig.~\ref{TP_Nt_g0} shows the $a\gamma_{0}^{*}$ dependence on $g_{0}$ for $N_{\tau}=2$ and $am_{0}=0.2$. A particularly interesting feature is that $\gamma_{0}$ displays a monotonically increasing $g_{0}$ behaviour within the considered coupling range. The growth of $a\gamma_{0}$ is anticipated, since larger values of $g_{0}$ imply that the thermoparticle states experience the effects of the medium more strongly. In Fig.~\ref{TP_am0} we plot $a\gamma_{0}^{*}$ for $N_{\tau}=2$ and $g_{0}=1.0$ at different values of $am_{0}$. One can see that as $am_{0}$ increases the damping parameter successively decreases. This is also physically consistent, since larger values of $am_{0}$ imply larger vacuum masses, and hence the corresponding thermoparticle states are less sensitive to in-medium effects. \\

\noindent
Overall, the results in this section provide strong numerical evidence that thermoparticles are the correct degrees of freedom with which finite-temperature perturbative calculations should be performed. Thermoparticle perturbation theory not only gives rise to precise predictions of finite-temperature lattice correlation function data, but the behaviour of the thermoparticle scattering states is physically consistent with the expected properties of the thermal medium.

\section{Conclusions} 
\label{concl}

In this work we continue the previous critiques of the standard approach to finite-temperature perturbation theory, and detail how the well-known issues of pinch singularities, infrared divergences, and poor convergence properties, are all related to the inconsistent thermal generalisation of the Gell-Mann-Low (GML) relation. The standard GML relation is based on free on-shell asymptotic scattering states, which in thermal equilibrium are known not to exist for interacting theories due to the presence of finite-temperature effects at large times, as implied by the Narnhofer-Requardt-Thirring theorem. To resolve these issues we instead propose a finite-temperature GML relation using thermoparticles as the thermal asymptotic states. Thermoparticles are damped but stable particle-like excitations, and thus carry information about the dynamics of the theory. A perturbative expansion based on this modified GML relation produces exactly the same Feynman diagrams as the standard approach, but now the propagators in these diagrams are those of the thermoparticle states. This implies that damping effects enter already at leading order, and this regularises both pinch singularities and infrared divergences by a physical effect, rather than for computational consistency purposes, as in the standard approach. Another important characteristic is that the damping of the thermal scattering states is model dependent, which reflects the fact that medium effects themselves are determined by the dynamics of the specific theory. We also demonstrate that the ultraviolet divergences appearing in the thermoparticle perturbative series are no more severe than in the vacuum theory, and hence the renormalisation of the theory at $T=0$ should be sufficient to guarantee finite perturbative predictions for any $T>0$. \\ 

\noindent
In the final part of this work we used numerical lattice simulations to test the validity of thermoparticle perturbation theory in massive $\phi^{4}$ theory. By performing lattice perturbation theory calculations of two-point correlation functions and comparing these with lattice data, we explicitly demonstrate that the predictions of the standard approach deviate significantly from the lattice data as the temperature is increased, whereas the thermoparticle approach gives rise to precise predictions at the 1\% accuracy level already at two-loop order. This is strong evidence that thermoparticles are indeed the correct degrees of freedom for performing consistent finite-temperature perturbative expansions, and that their use may have a significant impact on the predictions of various other observables, including thermodynamic potentials, screening masses, and transport coefficients. Although the focus of this study was scalar QFTs, these results have implications for the perturbative analysis of more complex theories such as QCD, where non-perturbative evidence of thermoparticles has already been found~\cite{Lowdon:2022xcl,Bala:2023iqu}.

\section*{Acknowledgements}
The authors acknowledge support by the Deutsche Forschungsgemeinschaft (DFG, German Research Foundation) through the Collaborative Research Center CRC-TR 211 ``Strong-interaction matter under extreme conditions'' -- Project No. 315477589-TRR 211.

\bibliographystyle{JHEP}

\bibliography{refs}

\end{document}